\documentclass[preprint,3p,12pt,sort&compress]{elsarticle}
\usepackage{mathrsfs}
\usepackage{amsmath}
\usepackage{stmaryrd}
\usepackage{bbding}
\usepackage{dcolumn}
\usepackage{graphicx}
\usepackage{amsfonts}
\usepackage{amssymb}
\usepackage{psfrag}
\usepackage{wrapfig}
\usepackage{subfigure}
\usepackage{makeidx}
\usepackage{bm}
\usepackage{epsf}
\usepackage{epsfig}
\usepackage{setspace}
\usepackage{graphicx}
\usepackage{epstopdf}
\usepackage{psfrag}
\usepackage{subfigure}
\usepackage{color}
\usepackage{enumerate}
\usepackage{verbatim} 
\usepackage{booktabs}
\usepackage[utf8]{inputenc}

\begin{document}

\title{A flux reconstruction kinetic scheme for the Boltzmann equation}
\author[KIT]{Tianbai Xiao}
\ead{tianbaixiao@gmail.com}
\address[KIT]{Karlsruhe Institute of Technology, Karlsruhe, Germany}

\begin{abstract}
    It is challenging to solve the Boltzmann equation accurately due to the extremely high dimensionality and nonlinearity.
    This paper addresses the idea and implementation of the first flux reconstruction method for high-order Boltzmann solutions.
    Based on the Lagrange interpolation and reconstruction, the kinetic upwind flux functions are solved simultaneously within physical and particle velocity space.
    The fast spectral method is incorporated to solve the full Boltzmann collision integral with a general collision kernel.
    The explicit singly diagonally implicit Runge-Kutta (ESDIRK) method is employed as time integrator and the stiffness of the collision term is smoothly overcome.
    Besides, we ensure the shock capturing property by introducing a self-adaptive artificial dissipation, which is derived naturally from the effective cell Knudsen number at the kinetic scale.
    As a result, the current flux reconstruction kinetic scheme can be universally applied in all flow regimes.
    Numerical experiments including wave propagation, normal shock structure, one-dimensional Riemann problem, Couette flow and lid-driven cavity will be presented to validate the scheme.
    The order of convergence of the current scheme is clearly identified.
    The capability for simulating cross-scale and non-equilibrium flow dynamics is demonstrated.
\end{abstract}


\maketitle

\section{Introduction}

The computational fluid dynamics (CFD) has been in voracious self-evolution in recent decades.
A highly visible direction is the development of high-order numerical methods.
In spite of the benefits from being intuitive, robust and easy for implementation, the traditional second-order methods have proven to
be insufficiently accurate under a comparable computational cost \cite{wang2013high}.
The high resolution and low dissipation inherited by high-order methods enable high-fidelity simulation of intricate flows in turbulence, acoustics, plasma physics, etc.
It is more complex to implement high-order methods and they are basically less robust than first- and second-order schemes.

High-order methods have been developed in the context of the finite difference (FD), finite volume (FV) and finite element (FE) formulations.
By extending the difference stencils, higher-order finite difference methods can be constructed and it is feasible to construct compact stencils \cite{lele1992compact}.
However, the straightforward extensions are restricted to problem domains with regular geometry only \cite{liszka1980finite}.
The finite volume methods can handle complex geometries in design, and a series of high-order extensions have been developed with regular and irregular geometries \cite{harten1987uniformly,abgrall1994essentially,liu1994weighted,jiang1996efficient}.
The reconstructions in FV methods are mostly based on cell-averaged values, resulting in non-compact stencils.

The thriving finite element methods provide an alternative to design high-order methods.
The discontinuous Galerkin (DG) method is arguably one of the most studied high-order FE algorithms \cite{cockburn2001runge,cockburn1998local,peraire2008compact,luo2010reconstructed}, which originates from the work on neutron transport problem by Reed and Hill \cite{reed1973triangular}.
The basic idea of the DG methods lies in the unified consideration of spatial discretization and spectral decomposition.
Within each element, the solutions are represented via polynomial basis functions and are allowed to be discontinuous across cell boundaries, which encourages the method to capture sharp wave structures that arise in fluid mechanics.
Thanks to the in-cell polynomials, it is straightforward to extend the DG methods to arbitrarily order of accuracy for smooth solutions.
As a special case of DG methods, the nodal DG scheme employs Lagrange polynomials as basis functions to interpolate solutions between distinct nodal points \cite{hesthaven2007nodal}.
Such idea is implemented similarly in another class of algorithms named the spectral difference (SD) methods \cite{kopriva1996conservative,liu2006spectral}, but based on the differential form of governing equations.

Huynh's work on the flux reconstruction (FR) approach provides profound insight into constructing high-order methods for any
advection-diffusion type equation \cite{huynh2007flux}.
It establishes a general framework, where many existing approaches such as the nodal DG and spectral difference methods can be understood as its particular cases.
Jameson used the FR formulation to prove that the SD method is uniformly stable in a norm of Sobolev type provided that the flux collocation points are placed at the zeros of the corresponding Legendre polynomial \cite{jameson2010proof}.
The essential connections between FR and DG methods have been analyzed in \cite{allaneau2011connections,de2014connections}.
A series of flux reconstruction methods have been developed correspondingly \cite{vincent2011new,castonguay2012new,asthana2015high,vandenhoeck2019implicit,li2020high}.
Specifically, Vincent et al. proposed a new class of energy stable flux reconstruction methods based on Huynh's approach, which is often referred as Vincent-Castonguay-Jameson-Huynh (VCJH) schemes \cite{vincent2011new}.
In what follows, we refer the terminology Flux Reconstruction corresponding to Vincent's formulation if unspecified.

Another hot topic in computational fluid dynamics research might go into the study of multi-scale and non-equilibrium flow dynamics.
As an example, the Boltzmann equation provides a statistical description of particle transports and collisions at the mesoscopic scale, i.e. the molecular mean free path and collision time.
The evolution of one-particle probability density function is followed within the phase space.
Compared to macroscopic fluid equations, the Boltzmann equation provides many more degrees of freedom
and thus can be used to describe both equilibrium and non-equilibrium systems.
Hilbert’s 6th problem \cite{hilbert1902mathematical} served as an intriguing beginning of trying to link the behaviors of an interacting many-particle system across different scales.
It has been shown since then that hydrodynamic equations can be
recovered from the asymptotic limits of the Boltzmann solutions \cite{chapman1990mathematical,grad1949kinetic}.

Continuous Efforts efforts have been devoted to the construction of numerical solvers for the Boltzmann and its related equations \cite{goldstein1989investigations,bobylev1997difference,mieussens2000discrete,filbet2010class,xu2010unified,xiao2017well,xiao2019unified}.
It is challenging to solve the Boltzmann equation accurately and efficiently due to the extremely high dimensionality and nonlinearity, and most of the solvers above are no more than second order accuracy.
The existing attempts on constructing high-order Boltzmann solvers are very limited.
Boscheri and Dimarco \cite{boscheri2020high} developed a class of central WENO implicit-explicit Runge Kutta schemes for the simplified BGK model of the Boltzmann equation.
Jaiswal et al. \cite{jaiswal2019discontinuous} and Su el al. \cite{su2020implicit} developed the discontinuous Galerkin methods for the Boltzmann equation.
To the best of the author's knowledge, no preliminary work has been done on developing the flux reconstruction method for the Boltzmann equation.

In this paper, a novel flux reconstruction kinetic scheme (FRKS) is presented for the Boltzmann equation.
Based on the Lagrange interpolation and reconstruction, the kinetic upwind flux functions are solved simultaneously within physical and particle velocity space.
The fast spectral method is incorporated into the FR framework to solve the full Boltzmann collision integral.
The explicit singly diagonally implicit Runge-Kutta (ESDIRK) method \cite{kennedy2016diagonally} is incorporated as numerical integrator and thus the stiffness of the collision operator in the continuum flow regime can be overcome.
We ensure the shock capturing property by introducing a self-adaptive artificial dissipation, which is derived from the effective cell Knudsen number at the kinetic scale.
As a result, the FRKS is able to capture the cross-scale flow dynamics where resolved and unresolved regions coexist inside a flow field.

The rest of this paper is organized as follows. 
Section \ref{sec:theory} is a brief introduction of the kinetic theory of gases. 
Section \ref{sec:algorithm} presents the formulation of the solution algorithm and its detailed implementation.
Section 4 includes numerical experiments to demonstrate the performance of the flux reconstruction kinetic scheme. 
The last section is the conclusion.

\section{Kinetic Theory}\label{sec:theory}

The gas kinetic theory describes the time-space evolution of particle distribution function $f(t,\mathbf x,\mathbf v)$. 
With a separate modeling of particle transport and collision processes, the Boltzmann equation of dilute monatomic gas in the absence of external force is
\begin{equation}
    \frac{\partial f}{\partial t} + \mathbf v \cdot \nabla_\mathbf x f = Q(f,f) = \int_{\mathbb{R}^{3}} \int_{\mathbb S^{2}}\left[f\left(\mathbf{v}^{\prime}\right) f\left(\mathbf{v}_{*}^{\prime}\right)-f(\mathbf{v}) f\left(\mathbf{v}_{*}\right)\right] \mathcal{B}(\cos \theta, g) d \mathbf \Omega d \mathbf{v}_{*},
\end{equation}
where $\{\mathbf v, \mathbf v_*\}$ are the pre-collision velocities of two classes of colliding particles, and $\{\mathbf v', \mathbf v_*'\}$
are the corresponding post-collision velocities.
The collision kernel $\mathcal{B}(\cos \theta, g)$ measures the probability of collisions in different directions, where $\theta$ is the deflection angle and $g = |\mathbf g| = |\mathbf v - \mathbf v_*|$ is the magnitude of relative pre-collision velocity.
The solid angle $\mathbf \Omega$ is the unit vector along the relative post-collision velocity $\mathbf v' - \mathbf v_*'$, and the deflection angle satisfies the relation $\theta=\mathbf \Omega \cdot \mathbf g / g$.
With the collision frequency defined as
\begin{equation}
    \nu(\mathbf v)=\int_{\mathbb{R}^{3}} \int_{\mathbb{S}^{2}} f\left(\mathbf v_{*} \right) \mathcal B\left(\cos \theta,g\right) d \mathbf \Omega d \mathbf v_{*},
\end{equation}
The Boltzmann collision integral can be written as a combination of gain and loss, i.e.,
\begin{equation}
    Q(f,f) = Q^+ + Q^- = \int_{\mathbb{R}^{3}} \int_{\mathbb S^{2}} f\left(\mathbf{v}^{\prime}\right) f\left(\mathbf{v}_{*}^{\prime}\right) \mathcal{B}(\cos \theta, g) d \mathbf \Omega d \mathbf{v}_{*} - \nu(\mathbf v) f(\mathbf v).
\end{equation}

A particle distribution function is related to unique macroscopic state.
The conservative flow variables can be obtained from the velocity moments of distribution function, i.e.
\begin{equation}
    \mathbf{W}(t, \mathbf{x})=\left(\begin{array}{c}
    \rho \\
    \rho \mathbf{V} \\
    \rho E
    \end{array}\right) := \int_{\mathbb R^3} f \psi d \mathbf{v},
\end{equation}
where $\psi=[1,\mathbf v,\mathbf v^2/2]^T$ is a vector of collision invariants satisfying $\int_{\mathbb R^3} Q(f,f) \psi d \mathbf v = 0$, and temperature is defined as
\begin{equation}
    \frac{3}{2} kT = \frac{1}{2n} \int (\mathbf v - \mathbf V)^2 f d\mathbf v,
\end{equation}
where $k$ is the Boltzmann constant and $n$ is the number density of gas.

Substituting the function $H=\int_{\mathbb R^3} \int_{\mathbb R^3} f \log f d\mathbf v d\mathbf x$ into the Boltzmann equation, we have
\begin{equation}
\begin{aligned}
    \frac{d H}{d t}&=-\int_{\mathbb{R}^{3}} \int_{\mathbb{R}^{3}}(\log f+1) \mathbf v \cdot \nabla_{\mathbf x} f d \mathbf v d \mathbf x+\int_{\mathbb{R}^{3}} \int_{\mathbb{R}^{3}}(\log f+1) Q(f,f) d \mathbf v d \mathbf x \\
    &= \int_{\mathbb{R}^{3}} \int_{\mathbb{R}^{3}} \log f Q(f,f) d \mathbf v d \mathbf x.
\end{aligned}
\end{equation}
From the H-theorem \cite{chapman1990mathematical}, we learn that the $H$ function is minimal only if $f$ is a Maxwellian,
\begin{equation}
    f=\mathcal M := \rho\left(\frac{m}{2\pi k T}\right)^{3/2} \exp(-\frac{m}{2kT} \left(\mathbf v - \mathbf V)^2 \right),
\end{equation}
where $m$ is the molecular mass.

\section{Solution Algorithm}\label{sec:algorithm}

\subsection{Formulation}

Considering the domain $\boldsymbol{\Omega}$ with $N$ non-overlapping cells
\begin{equation}
    \boldsymbol{\Omega}=\bigcup_{i=1}^{N} \boldsymbol{\Omega}_{i}, \quad \bigcap_{i=1}^{N} \boldsymbol{\Omega}_{i}=\emptyset,
\end{equation}
we represent the solution of the Boltzmann equation with piecewise polynomials.
Within each element $\boldsymbol{\Omega}_{i}$, the particle distribution function is approximated by a polynomial of degree $m$ denoted $f_i \simeq f_i^\delta(t,\mathbf x,\mathbf v)$, and the corresponding flux function is approximated of degree $m+1$, i.e. $F_i \simeq F_i^\delta(t,\mathbf x,\mathbf v)$.
Therefore, the total approximate solutions are
\begin{equation}
    f^{\delta}=\bigoplus_{i=1}^{N} f_{i}^{\delta} \approx f, \quad F^{\delta}=\bigoplus_{i=1}^{N} F_{i}^{\delta} \approx F.
\end{equation}

For convenience, a standard coordinate can be introduce locally as $\boldsymbol{\Omega}_{S}=\{\mathbf r | \mathbf r \in [-1,1]^3\}$.
The transformation of coordinates is made by the mapping
\begin{equation}
    \mathbf r = \Gamma_i(\mathbf x) = \left[\begin{array}{c}
2\left(\frac{x-x_{i-1/2}}{x_{i+1/2}-x_{i-1/2}}\right)-1 \\
2\left(\frac{y-y_{j-1/2}}{y_{j+1/2}-y_{j-1/2}}\right)-1 \\
2\left(\frac{z-z_{k-1/2}}{z_{k+1/2}-z_{k-1/2}}\right)-1 \\
\end{array}\right].
\end{equation}
Here we take structured mesh for illustration, while the mapping in unstructured mesh can be found in \cite{castonguay2012new,witherden2014pyfr}.
And thus the Boltzmann equation in the local coordinate system becomes
\begin{equation}
    \frac{\partial \hat{f}^{\delta}}{\partial t}=-\nabla_{\mathbf r} \cdot \hat{\mathbf F}^{\delta}+\hat Q^\delta,
    \label{eqn:fr update}
\end{equation}
where $\hat{\mathbf F}^\delta$ and $\hat Q^\delta$ are the numerical flux and collision term respectively.

\subsection{Flux}

\subsubsection{Discontinuous flux}\label{sec:discontinuous flux}

In the flux reconstruction method, the solution is approximated by piecewise polynomials.
For brevity, let us take one-dimensional geometry as example, while the extension to multi-dimensional case is straightforward via tensorization.
We define the following Lagrange polynomials of degree $m$
\begin{equation}
    l_{p}=\prod_{q=0, q \neq p}^{m}\left(\frac{r-r_{q}}{r_{p}-r_{q}}\right),
\end{equation}
and the particle distribution function can be represented on the basis of $m+1$ solution points
\begin{equation}
    \hat{f}^{\delta}=\sum_{p=0}^{m} \hat{f}_{p}^{\delta} l_{p}.
\end{equation}

For the Boltzmann equation, the flux function is defined as
\begin{equation}
    F(t,x,v)=vf(t,x,v),
\end{equation}
and thus the local flux function is
\begin{equation}
    \hat{F}(t, r, v)=\frac{F \left(t,\Gamma_{i}^{-1}(r), v\right)}{J_{i}},
    \label{eqn:local direct flux}
\end{equation}
where $J_i=(x_{i+1/2}-x_{i-1/2})/2$ is the Jacobian.
Therefore, the flux polynomials can be constructed as
\begin{equation}
    \hat{F}^{\delta D}=\sum_{p=0}^{m} \hat{F}_{p}^{\delta D} l_{p},
    \label{eqn:discontinuou flux polynomial}
\end{equation}
where $\hat{F}_{p}^{\delta D}$ is the flux calculated by Eq.(\ref{eqn:local direct flux}) at solution point $r_p$.
The notation $\delta D$ implies that such a flux is basically discontinuous since it is derived directly from piecewise discontinuous solutions of $\hat f^\delta$.

\subsubsection{Interactive flux}

The discontinuou flux polynomials in Eq.(\ref{eqn:local direct flux}) is of the same degree of freedom $m$ as solutions, which doesn't meet the accuracy requirement.
Besides, it doesn't take the information from adjacent cells into consideration and can by no means deal with gas-surface interactions.
A natural idea is to introduce a degree $m+1$ correction flux to the approximate transformed discontinuous flux, i.e.
\begin{equation}
    \hat{F}^{\delta} = \hat{F}^{\delta D} + \hat{F}^{\delta C}.
\end{equation}
The total flux is expected to equal the correct interactive fluxes at cell boundaries, and to preserve a similar in-cell profile of discontinuous flux.
A feasible approach, as proposed in \cite{huynh2007flux}, is to introduce two symmetric auxiliary functions $\{ g_L, g_R \}$, which satisfy
\begin{equation}
\begin{aligned}
    & g_L(r) = g_R(-r), \\
    &g_L(-1)=1, \ g_R(-1)=0, \\
    &g_L(1)=0, \ g_R(1)=1. \\
\end{aligned}
\end{equation}
The corresponding correction flux can be reconstructed as
\begin{equation}
    \hat{F}^{\delta C} = (\hat{F}^{\delta I}_L - \hat{F}^{\delta D}_L) g_L + (\hat{F}^{\delta I}_R - \hat{F}^{\delta D}_R) g_R.
\end{equation}
Here $\{ \hat{F}^{\delta D}_L,\hat{F}^{\delta D}_R \}$ are the reconstructed discontinuous fluxes from the polynomial representation at the left and right boundary of element,
and $\{ \hat{F}^{\delta I}_L,\hat{F}^{\delta I}_R \}$ are the interactive fluxes at the boundaries.
In the Boltzmann equation, we can clearly identify the flight directions of particle transports, and the corresponding upwind flux can be evaluated as
\begin{equation}
\begin{aligned}
    \hat{F}^{\delta I}_{i,L} = \hat{F}^{\delta D}_{i-1,R} H(v) + \hat{F}^{\delta D}_{i,L} H(1-v),\\
    \hat{F}^{\delta I}_{i,R} = \hat{F}^{\delta D}_{i,R} H(v) + \hat{F}^{\delta D}_{i+1,L} H(1-v),
\end{aligned}
\end{equation}
where $H(x)$ is the heaviside step function.

\subsubsection{Total flux}

Given the total flux $\hat{F}^\delta$, the its derivatives can be expressed as
\begin{equation}
\begin{aligned}
    \frac{\partial \hat{F}^{\delta}}{\partial r} = \frac{\partial \hat{F}^{\delta D}}{\partial r} + \frac{\partial \hat{F}^{\delta C}}{\partial r}. \\
\end{aligned}
\end{equation}
It can be evaluated by calculating the divergences of the Lagrange polynomials and the correction functions at each solution point $r_p$, i.e.
\begin{equation}
    \frac{\partial \hat{F}^{\delta}}{\partial r}\left(r_{p}\right)=\sum_{q=0}^{m} \hat{F}_{q}^{\delta D} \frac{\mathrm{d} l_{q}}{\mathrm{~d} r}\left(r_{p}\right)+\left(\hat{f}_{L}^{\delta I}-\hat{f}_{L}^{\delta D}\right) \frac{\mathrm{d} g_{L}}{\mathrm{~d} r}\left(r_{p}\right)+\left(\hat{f}_{R}^{\delta I}-\hat{f}_{R}^{\delta D}\right) \frac{\mathrm{d} g_{R}}{\mathrm{~d} r}\left(r_{p}\right).
\end{equation}

\subsection{Collision}

It is challenging to solve the Boltzmann collision integral due to the extremely high dimensionality and nonlinearity.
The numerical Boltzmann solvers are pioneered by Goldstein et al. \cite{goldstein1989investigations}.
The early approaches solves the fivefold integral with Eulerian grid points and interpolations \cite{sone1989temperature,ohwada1993structure}.
Given the two-body collision model, the computational cost of these methods are of $O(N^6)$, where $N$ is the number of discrete velocity points in one direction, and only half order of convergence is realized.

Another class of methods solves the Boltzmann equation with the Fourier transform.
Bobylev made a preliminary attempt of such method for the Maxwell molecules in a homogeneous flow field \cite{bobylev1988theory}.
In 2006, Mouhot and Pareschi proposed a fast spectral method based on the Carleman-type Boltzmann equation \cite{carleman1944integrale}, with the spectral accuracy and the computational cost of $O(M^2N^3\log N)$ \cite{mouhot2006fast}.
Here $M$ is the number of grid points for discretizing polar angles, which is much smaller than the number of velocity grids $N$ in each direction.
The advantageous efficiency and efficiency guarantee its dominance in wide applications \cite{wu2013deterministic,gamba2017fast,jaiswal2019discontinuous,su2020implicit,xiao2020velocity}. 
Here, we briefly introduce the basic idea of this method.

The Carleman-type Boltzmann equation can be obtained with the following transformations,
\begin{equation}
\begin{aligned}
Q(f,f) &=\int_{\mathbb{R}^{3}} \int_{\mathbb{S}^{2}} \Theta g\left[f\left(\mathbf{v}^{\prime}\right) f\left(\mathbf{v}_{*}^{\prime}\right)-f(\mathbf{v}) f\left(\mathbf{v}_{*}\right)\right] d \mathbf \Omega d \mathbf{v}_{*} \\
&=\int_{\mathbb{R}^{3}} \int_{\mathbb{S}^{2}} \Theta g\left[f\left(\mathbf{v}+\frac{g \Omega-\mathbf{g}}{2}\right) f\left(\mathbf{v}_{*}-\frac{g \Omega-\mathbf{g}}{2}\right)-f(\mathbf{v}) f\left(\mathbf{v}_{*}\right)\right] d \Omega d \mathbf{v}_{*} \\
&=2 \int_{\mathbb{R}^{3}} \int_{\mathbb{R}^{3}} \Theta \delta\left(2 \mathbf{y} \cdot \mathbf{g}+\mathbf{y}^{2}\right)\left[f\left(\mathbf{v}+\frac{\mathbf{y}}{2}\right) f\left(\mathbf{v}_{1}-\frac{\mathbf{y}}{2}\right)-f(\mathbf{v}) f\left(\mathbf{v}_{*}\right)\right] d \mathbf{y} d \mathbf{v}_{*} \\
&=4 \int_{\mathbb{R}^{3}} \int_{\mathbb{R}^{3}} \Theta \delta\left(\mathbf{y} \cdot \mathbf{g}+\mathbf{y}^{2}\right)\left[f(\mathbf{v}+\mathbf{y}) f\left(\mathbf{v}_{*}-\mathbf{y}\right)-f(\mathbf{v}) f\left(\mathbf{v}_{*}\right)\right] d \mathbf{y} d \mathbf{v}_{*} \\
&=4 \int_{\mathbb{R}^{3}} \int_{\mathbb{R}^{3}} \Theta \delta(\mathbf{y} \cdot \mathbf{z})[f(\mathbf{v}+\mathbf{y}) f(\mathbf{v}+\mathbf{z})-f(\mathbf{v}) f(\mathbf{v}+\mathbf{y}+\mathbf{z})] d \mathbf{y} d \mathbf{z},
\end{aligned}
\end{equation}
where $\Theta=\mathcal B/g$ is the differential cross section, $\mathbf y=(g\mathbf \Omega - \mathbf g)/2$ and $\mathbf z = \mathbf v_* - \mathbf v - \mathbf y=-\mathbf g - \mathbf y$.

As proposed in \cite{wu2013deterministic}, a general collision kernel can be represented as
\begin{equation}
    \mathcal{B}=C_{\alpha, \gamma}^{\prime \prime} \sin ^{\alpha+\gamma-1}\left(\frac{\theta}{2}\right) \cos ^{-\gamma}\left(\frac{\theta}{2}\right) g^{\alpha},
\end{equation}
with
\begin{equation}
    C_{\alpha, \gamma}^{\prime \prime}=\frac{\Gamma[(7+\alpha) / 2]}{6 \Gamma[(3+\alpha+\gamma) / 2] \Gamma(2-\gamma / 2)} C_{\alpha},
\end{equation}
where $C_\alpha$ is the factor given in \cite{bird1994molecular}.
For commonly-used molecular models, the above equation can be simplified.
For example, it reduces to 
\begin{equation}
    \mathcal B = \frac{1}{4}gd^2
\end{equation}
for the hard-sphere (HS) model.

In the fast spectral method, the particle distribution function is discretized with $N=[N_u,N_v,N_w]^T$ quadrature points and periodized in a truncated domain $\mathcal D = [-L, L]^3$.
For the particle distribution functions in a standard element $\boldsymbol \Omega_S$ of the flux reconstruction scheme, the Fourier series can be constructed as,
\begin{equation}
\begin{aligned}
&\hat f^\delta(t,\mathbf r,\mathbf{v})=\sum_{k=-N / 2}^{N / 2-1} \hat{f}^\delta_{k}(t,\mathbf r) \exp \left(i \xi_{k} \cdot \mathbf{v}\right), \\
&\hat{f}^\delta_{k}=\frac{1}{(2 L)^{3}} \int_{D_{L}} \hat f^\delta(t,\mathbf r,\mathbf{v}) \exp \left(-i \xi_{k} \cdot \mathbf{v}\right) d \mathbf{v},
\end{aligned}
\end{equation}
where $i$ is the imaginary unit and $\xi_k=k\pi/L$ is the frequency component.
Similarly, the collision operator can be expanded as
\begin{equation}
    \hat{Q}^\delta_{k}=\sum_{l, m=-N / 2,(l+m=k)}^{N / 2-1} \hat{f}_{l}^\delta \hat{f}_{m}^\delta [\beta(l, m)-\beta(m, m)],
\end{equation}
where $l=[l_x,l_y,l_z]^T$ and $m=[m_x,m_y,m_z]^T$.
The kernel mode is
\begin{equation}
\begin{aligned}
\beta(l, m)&=4 \int_{\mathbb{R}^{3}} \int_{\mathbb{R}^{3}} \delta(\mathbf{y} \cdot \mathbf{z}) \Theta \exp \left(i \xi_{l} \cdot \mathbf{y}+i \xi_{m} \cdot \mathbf{z}\right) d \mathbf{y} d \mathbf{z} \\
&=\iint \delta\left(\mathbf{e} \cdot \mathbf{e}^{\prime}\right) \Theta\left[\int_{-R}^{R} \rho \exp \left(i \rho \xi_{l} \cdot \mathbf{e}\right) d \rho\right]\left[\int_{-R}^{R} \rho^{\prime} \exp \left(i \rho^{\prime} \xi_{m} \cdot \mathbf{e}^{\prime}\right) d \rho^{\prime}\right] d \mathbf{e} d \mathbf{e}^{\prime} \\
&=4 \int\left[\int_{0}^{R} \rho \cos \left(\rho \xi_{l} \cdot \mathbf{e}\right) d \rho\right]\left[\int \delta\left(\mathbf{e} \cdot \mathbf{e}^{\prime}\right) \int_{0}^{R} \rho^{\prime} \Theta \cos \left(\rho^{\prime} \xi_{m} \cdot \mathbf{e}^{\prime}\right) d \rho^{\prime} d \mathbf{e}^{\prime}\right] d \mathbf{e}
\end{aligned}
\end{equation}
with $\mathbf e$ and $\mathbf e'$ being the vectors on the unit sphere $\mathbb S^2$.
The splitting of $l$ and $m$ kernel modes can be achieved by numerical quadrature.
Afterwards the convolution from the Boltzmann collision integral can be evaluated efficiently with the fast Fourier transform in the frequency domain.
To avoid tedious repetition, we refer \cite{mouhot2006fast} for more details of the fast spectral method.

\subsection{Integrator}

After finishing the evaluations of fluxes and collision terms, we get the time derivatives of particle distribution function at the solution points $\{\mathbf r_i,\mathbf v_j\}$ from Eq.(\ref{eqn:fr update}), i.e.
\begin{equation}
    \frac{\partial \hat{f}_{i,j}^{\delta}}{\partial t}=
    -\nabla_{\mathbf r} \cdot \hat{\mathbf F}_{i,j}^{\delta} + \hat Q_{i,j}^\delta = \hat{\mathcal C}^\delta_{i,j},
\end{equation}
where $\hat{\mathcal C}_{i,j}^\delta={\mathcal C}_j(\hat f^\delta_{i})$ denotes a combination of flux and collision operators.
Note that $\hat Q_{i,j}^\delta$ can become stiff in the continuum limit, when the particle distribution function is close to the Maxwellian \cite{filbet2010class}.
To circumvent the CFL restriction, an appropriate time integrator needs to be chosen in hope that it is efficient and A- or L-stable for stiff and oscillatory problems.

A prevailing family of integration methods for stiff differential equations is the backward differentiation formula (BDF) thanks to its ease of implementation \cite{curtiss1952integration}.
As linear multi-step methods, the BDF methods with an order greater than two cannot be A-stable.
In spite of the attempts on constructing higher-order A-stable methods by introducing additional stages \cite{carpenter2003efficiency}, these methods haven't been proven to be universally effective and thus the most commonly used method is BDF-2.

An alternative integrator is the multi-stage implicit Runge–Kutta (IRK) methods \cite{jameson2017evaluation}.
In the original IRK methods, a fully coupled nonlinear system needs to be solved at each step or each stage.
To reduce the computational complexity, the diagonally implicit Runge–Kutta (DIRK) and singly diagonally implicit Runge–Kutta (SDIRK) methods have been proposed \cite{kennedy2016diagonally}.
As a further simplification, the explicit singly diagonally implicit Runge-Kutta (ESDIRK) method employs an explicit first step and thus reduces the degree of the nonlinear systems from SDIRK by one.
The comparisons from compressible Navier-Stokes equations indicated that the ESDIRK methods are more efficient than the BDF methods \cite{bijl2002implicit,wang2020comparison}.

In this paper, the A-L stable ESDIRK method is employed to construct the flux reconstruction scheme.
We provide a brief introduction of this method, while the comprehensive numerical implementation can be found in \cite{kennedy2003additive}.
The SDIRK method with $s$ stages can be written into the following form,
\begin{equation}
\begin{aligned}
    & t^p = t^n + c_p \Delta t^n, \\
    & \hat{f}^{p}_{i,j} = \hat{f}^{n}_{i,j}+\Delta t^{n} \sum_{q=1}^{p} a_{pq} \mathcal{C}_j\left(\hat{f}_{i}^q\right), \ p=1, \ldots, s, \\
    & \hat{f}^{n+1}_{i,j} = \hat{f}^{n}+\Delta t^{n} \sum_{p=1}^{s} b_{p} \mathcal{C}_j\left(\hat{f}^p_{i}\right).
\end{aligned}
\end{equation}
For the ESDIRK method, the first step is explicit and thus $a_{11}=0$.
The Butcher tableau of SDIRK and ESDIRK methods is presented in Table \ref{tab:butcher},
where the $s$ or $s-1$ nonlinear equation systems to be solved are clearly identified.
\begin{table}[htbp]
    \centering
    \caption{Butcher tableau of SDIRK (left) and ESDIRK (right) methods.}
    \begin{tabular}{c|cccc}
    $c_{1}$ & $a_{11}$ & 0 & $\cdots$ & 0 \\
    $c_{2}$ & $a_{21}$ & $a_{22}$ & $\cdots$ & 0 \\
    $\vdots$ & $\vdots$ & $\vdots$ & $\ddots$ & $\vdots$ \\
    $c_{s}$ & $a_{s1}$ & $a_{s2}$ & $\cdots$ & $a_{ss}$ \\
    \hline & $b_{1}$ & $b_{2}$ & $\cdots$ & $b_{s}$ \\
    \end{tabular}
    \quad
    \begin{tabular}{c|cccc}
    $0$ & $0$ & 0 & $\cdots$ & 0 \\
    $c_{2}$ & $a_{21}$ & $a_{22}$ & $\cdots$ & 0 \\
    $\vdots$ & $\vdots$ & $\vdots$ & $\ddots$ & $\vdots$ \\
    $c_{s}$ & $a_{s1}$ & $a_{s2}$ & $\cdots$ & $a_{ss}$ \\
    \hline & $b_{1}$ & $b_{2}$ & $\cdots$ & $b_{s}$ \\
    \end{tabular}
    \label{tab:butcher}
\end{table}

The advantage of DIRK-type integrators is that the computation of the stage vectors is decoupled. 
With $N_\mathbf r$ physical solution points and $N_\mathbf v$ velocity grid points, instead of solving one nonlinear system with $s\times N_\mathbf r \times N_\mathbf v$ unknowns, $s$ nonlinear systems with $N_\mathbf r \times N_\mathbf v$ unknowns are solved in practice. 
The solution algorithm at each implicit stage can be written as
\begin{equation}
    \frac{\hat{f}^{p}_{i,j}-\hat{f}^{n}_{i,j}}{a_{p p} \Delta t^{n}}=\mathcal{C}_j\left(\hat{f}_i^{p}\right)+\frac{1}{a_{p p}} \sum_{q=1}^{p-1} a_{p q} \mathcal{C}_j\left(\hat{f}^{q}_i\right),
\end{equation}
where the stage vectors and derivatives can be obtained via
\begin{equation}
\begin{aligned}
    & {s}^{p}=\hat{f}^{n}_{i,j}+\Delta t^{n} \sum_{q=1}^{p-1} a_{pq} \mathcal{C}_j\left(\hat{f}_i^{q}\right), \\
    & \mathcal{C}_j\left(\hat{f}_{i}^{p}\right)=\frac{1}{a_{p p} \Delta t^{n}}\left(\hat{f}^{p}_{i,j}-{s}^{p}\right).
\end{aligned}
\end{equation}

\subsection{Artificial dissipation}\label{sec:artificial dissipation}

Robust shock capturing is the critical factor for evaluating high-order methods in hyperbolic conservation laws.
In the vicinity of discontinuities in a self-evolving flow field, oscillations tend to appear due to the Gibbs
phenomenon and lead spurious or unstable solutions.
Given the less dissipation by nature, such effects are usually more severe for higher-order methods.
For the Boltzmann equation, the shock capturing is not as strongly desired since the shock structures can be resolved at particle mean free path level.
Here we still consider the handling of this issue as we expect to design a universal approach that can be applied in both resolved and unresolved regions.
We show that the high dimensional information from the Boltzmann equation can be extracted to inject more physically consistent artificial dissipation.

The issue of introducing artificial dissipation into high-order methods has been around for a long time.
The basic ideas can be categorized as follows.

\textbf{Limiting:} The idea is to co-opt the slope or flux limiters from finite volume methods based on certain rules, e.g. the total variation diminishing (TVD) or total variation bounded (TVB) principle. 
Preliminary work has been done in the context of discontinuous Galerkin methods \cite{cockburn1989tvb,krivodonova2004shock,krivodonova2007limiters}. 
In principle, such methods smear the discontinuous or sharp solutions across several adjacent cells, which significantly diminishes the significance of introducing solution points inside elements.
Besides, a naive usage of limiters can easily lead to descending order of accuracy around local extrema.

\textbf{Artificial viscosity:} An alternative way is to introduce artificial dissipative term around discontinuous regions.
The pioneer work was done in the Jameson-Schmidt-Turkel (JST) schemes \cite{jameson1981numerical}.
The idea has been implemented in the discontinuous Galerkin \cite{persson2006sub} and spectral difference methods \cite{premasuthan2014computation}.
The artifacts are expected to vanish in smooth regions and therefore the artificial viscosity coefficients in front of the even-order derivatives need to be solution or grid dependent \cite{cook2004high}.
It is extremely hard to set up a viscosity that can be universally applied to different equations or geometries.
Also, the design of boundary conditions is ambiguous for the artificial viscosity.

\textbf{Filtering:} Filters are commonly used in Galerkin-type methods \cite{flad2016simulation,frank2016convergence}.
The idea is to damp the high-order coefficients of the polynomials to eliminate high-frequency oscillations.
For the finite element type methods, the solution inside each element is a polynomial in essence, and can be expressed equivalently
with an orthogonal polynomial basis of the same degree \cite{sheshadri2016analysis}.
Different filter functions can be constructed and applied to the orthogonal polynomials, e.g. the $L^1$, $L^2$ and the exponential filters \cite{boyd2001chebyshev}.
The filtering plays basically as a separate step in the solution algorithm and is easy for implementation.
However, if the filter is applied everywhere as limiter in the domain, its strength needs to be very carefully chosen so that it doesn't destroy the key solution structure while mitigating the Gibbs phenomenon.
As a result, it is more often used locally with a detector of trouble cells \cite{vuik2014multiwavelet}.

As we stand on top of the Boltzmann equation, it provides us a different point of view to construct the artificial dissipation from the underlying kinetic physics.
Let us introduce the following dimensionless variables
\begin{equation}
    \tilde{\mathbf x} = \frac{\mathbf x}{L_0},  \ \tilde t = \frac{t}{L_0/V_0}, \ \tilde{\mathbf v} = \frac{\mathbf v}{V_0}, \ \tilde f = \frac{f}{n_0 V_0^3}
\end{equation}
where $V_0=\sqrt{2kT_0/m}$ is the most probable molecular speed,
and the Boltzmann equation becomes
\begin{equation}
    \frac{\partial \tilde f}{\partial \tilde t} + \tilde{\mathbf v} \cdot \nabla_{\tilde {\mathbf x}} {\tilde f} = \frac{1}{\mathrm{Kn}}Q(\tilde f, \tilde f).
\end{equation}
The Knudsen number is defined as
\begin{equation}
    \mathrm{Kn}=\frac{V_0}{L_0 \nu_0}=\frac{\ell_0}{L_0},
\end{equation}
where $\ell_0$ and $\nu_0$ are the molecular mean free path and mean collision frequency in the reference state.
For brevity, we drop the tilde notation to denote dimensionless variables henceforth.

As calculated in \cite{chang1956theory}, the thickness of a weak shock wave is around 10 molecular mean free paths, and is therefore of $O(10L_0\mathrm{Kn})$.
When the shock is resolved by the cell resolution, the Boltzmann equation is able to recover the physical solution profile.
Instead, if the fluid dynamics is solved at a coarser level, the shock becomes under resolved and thus performs as a discontinuity.
The sampling theorem indicates the best numerical solutions that can be captured under certain numerical resolution.
In this case, the physical shock thickness will be replaced by the numerical one anyway, where the finest discontinuity length equals the distance between two solution points.
An effective numerical dissipation can be introduce following this principle.

We modify the dimensionless Boltzmann equation as follows
\begin{equation}
    \frac{\partial f}{\partial t} + {\mathbf v} \cdot \nabla_{ {\mathbf x}} { f} = \frac{1}{\mathrm{Kn}_c}Q( f, f),
\end{equation}
where a cell Knudsen number is introduce instead of the original one.
The definition is given by
\begin{equation}
    \mathrm{Kn}_c = \mathrm{Kn} + \frac{\Delta x_m}{L_c}.
\end{equation}
Here $\Delta_m$ denotes the minimal distance between two adjacent solution points with polynomials of degree $m$.
A characteristic length scale of local cell is introduced based on the gradient,
\begin{equation}
    L_c = \frac{\phi}{\nabla_\mathbf x \phi}, 
\end{equation}
where $\phi$ is a physical quantity of interest.
Here we choose pressure as criterion of gradient, 
\begin{equation}
    p = \frac{1}{3} \int_{\mathcal R^3} (\mathbf v - \mathbf V) f d\mathbf v,
\end{equation}
and the evaluation of derivatives is conducted the same way as section \ref{sec:discontinuous flux}.

The modified Knudsen number is related to an augmented viscosity.
Let us define the symmetric linearized operator first,
\begin{equation}
    \mathcal L_{g}(f):=Q(g, f)+Q(f, g),
\end{equation}
where $g$ is another class of particle distribution functions.
Considering a small Knudsen number $\mathrm{Kn}_c=\varepsilon$, we can apply the Chapman-Enskog expansion to approximate the particle distribution function \cite{chapman1990mathematical},
\begin{equation}
    f \simeq f_\varepsilon = \mathcal M + \sum_{n=1}^\infty \varepsilon^n g_n,
\end{equation}
where $g_n \in \mathcal R(\mathcal L_\mathcal M)$.
As proved in \cite{saint2009hydrodynamic}, $\mathcal L_\mathcal M$ is self-adjoint with respect to $(f, g)_{\mathcal{M}}=\int_{\mathbb{R}^{3}} f(v) g(v)/\mathcal{M}(v) d v$.

Let $\mathcal P$ denote the projection $\mathcal N(\mathcal L_\mathcal M)$. 
Then applying $\mathcal P$ to the Boltzmann equation leads
\begin{equation}
    \mathcal P\left(\frac{\partial f}{\partial t}+\mathbf v \cdot \nabla_{\mathbf x} f\right)=\frac{1}{\varepsilon} \mathcal P \left(Q(f, f)\right).
\end{equation}
It is noticeable that $(Q(f, f), \psi)_{\mathcal{M}}=(Q(f, f), \psi \mathcal{M})_{\mathcal{M}}=0$ holds, so $Q(f, f) \in \mathcal{N}\left(\mathcal L_{\mathcal{M}}\right)^{\perp}=\mathcal{R}\left(\mathcal L_{\mathcal{M}}\right)$ implies
\begin{equation}
    \mathcal P\left({\partial_t f}+\mathbf v \cdot \nabla_{\mathbf x} f\right)=0.
\end{equation}
Inserting the Chapman-Enskog expansion into the equation above yields
\begin{equation}
    \mathcal P\left(\partial_{t}+\mathbf v \cdot \nabla_{\mathbf x}\right) \mathcal{M}=-\mathcal P\left(\partial_{t}+\mathbf v \cdot \nabla_{\mathbf x}\right)\left(\varepsilon g_{1}+\varepsilon^{2} g_{2}+\ldots\right).
\end{equation}
By matching the coefficients on the terms of order $\varepsilon$, we come to
\begin{equation}
    \left(\partial_{t}+\mathbf v \cdot \nabla_{\mathbf x}\right) \mathcal{M}=\mathcal L_{\mathcal{M}}\left(g_{1}\right).
\end{equation}
This equation has solution only if $\mathcal L_{\mathcal{M}}\left(g_{1}\right) \in \mathcal R(\mathcal L_\mathcal M)$.
Enforcing the projection onto $\mathcal R(\mathcal L_\mathcal M)$, and making use of the invertibility of $\mathcal L_\mathcal M$,
we get
\begin{equation}
    \begin{aligned}
    \mathcal P\left(\partial_{t}+\mathbf v \cdot \nabla_{\mathbf x}\right) \mathcal{M} &=-\varepsilon \mathcal P\left(\partial_{t}+\mathbf v \cdot \nabla_{\mathbf x}\right) \mathcal L_{\mathcal{M}}^{-1}(I-\mathcal P)\left(\partial_{t}+\mathbf v \cdot \nabla_{\mathbf x}\right) \mathcal{M} \\
    &=-\varepsilon \mathcal P\left(\mathbf v \cdot \nabla_{\mathbf x}\right) L_{\mathcal{M}}^{-1}(I-\mathcal P)\left(\mathbf v \cdot \nabla_{\mathbf x}\right) \mathcal{M},
    \end{aligned}
\end{equation}
which is the compact form of the Navier-Stokes equations.
As is shown, the augmented Knudsen number plays an equivalent role as artificial viscosity at the Navier-Stokes level.

\subsection{Summary}

\begin{figure}
    \centering
    \includegraphics[width=0.6\textwidth]{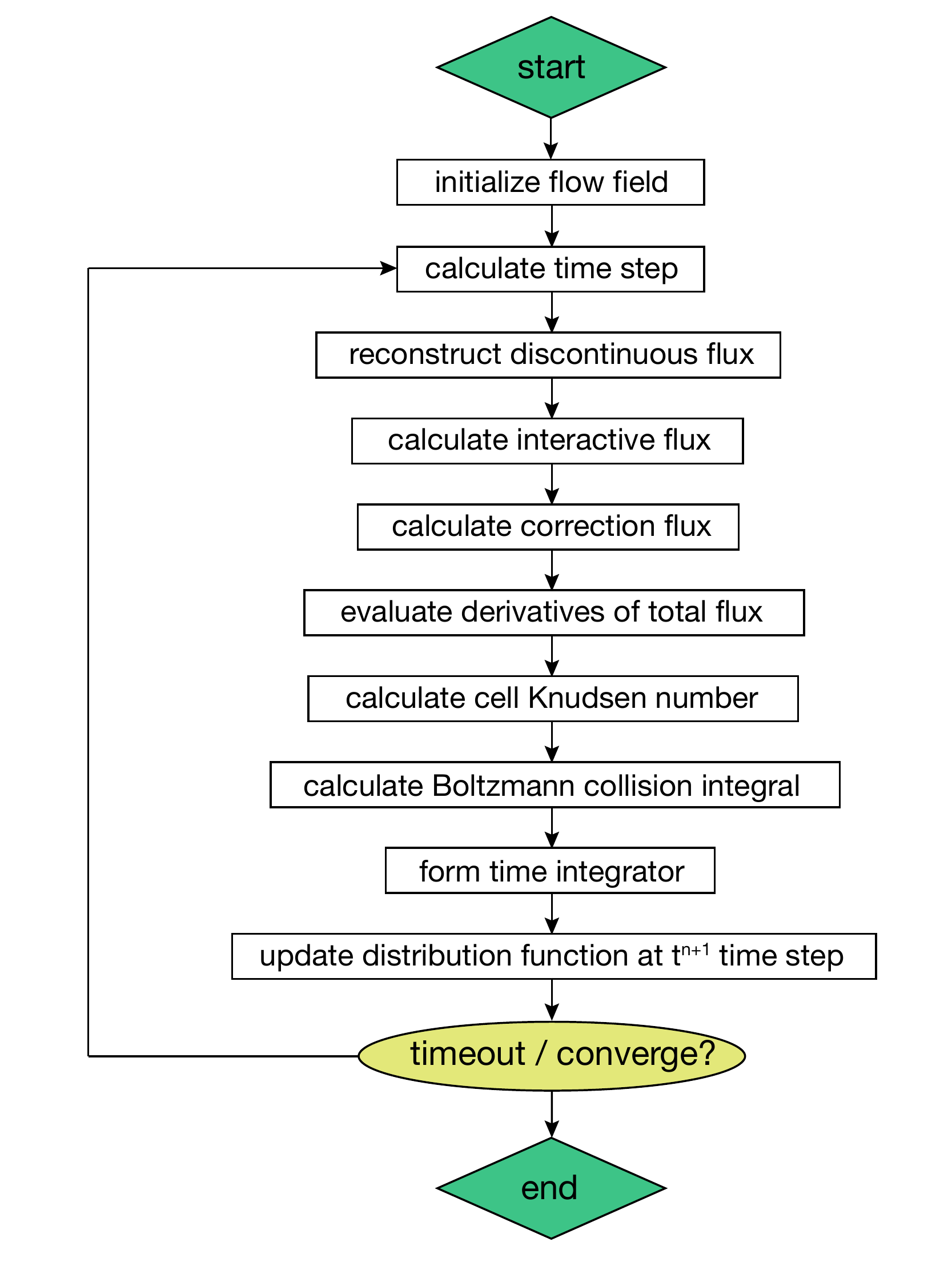}
    \caption{Flowchart of solution algorithm.}
    \label{fig:flowchart}
\end{figure}

The flowchart of the current solution algorithm is summarized in Fig. \ref{fig:flowchart}.

\section{Numerical Experiments}

In this section, we will conduct numerical experiments to validate the current scheme. 
In order to demonstrate the cross-scale computing capability of the algorithm, the results at different degrees of gas rarefaction are presented.
As introduced in section \ref{sec:artificial dissipation}, the dimensionless variables are used in all the numerical simulations.

\subsection{Wave propagation}

First we study the order of accuracy of the flux reconstruction kinetic scheme.
The propagation of an one-dimensional traveling wave is used as the test case.
The initial particle distribution function is set as Maxwellian in correspondence with the following macroscopic variables
\begin{equation*}
\left[\begin{array}{c}
\rho \\
U \\
V \\
W \\
p \\
\end{array}\right]_{{t=0}}=\left[\begin{array}{c}
1+\alpha \sin(2\pi x) \\
1  \\
0 \\
0 \\
0.5 \\
\end{array}\right],
\end{equation*}
and the detailed computational setup is presented in Table \ref{tab:wave}.
\begin{table}[htbp]
	\caption{Computational setup of wave propagation problem.} 
	\centering
	\begin{tabular}{lllllll} 
		\hline
		$t$ & $x$ & $N_x$ & Polynomial & Degree & Points & Correction \\ 
		\hline 
		$(0,1]$ & $[0,1]$ & $[5,40]$ & Lagrange & $[2,3]$ & Legendre & Radau \\ 
		\hline 
		$\mathbf v$ & $N_u$ & $N_v$ & $N_w$ & Quadrature & Kn & $\alpha$  \\ 
		\hline 
		$[-8,8]^3$ & 80 & 28 & 28 & Rectangular & $[0.0001,0.1]$ & 0.1  \\ 
		\hline
		Integrator & Boundary & CFL & \\
		\hline
		ESDIRK & Periodic & 0.1 & \\
		\hline
	\end{tabular} 
	\label{tab:wave}
\end{table}

\begin{figure}
    \centering
    \includegraphics[width=0.6\textwidth]{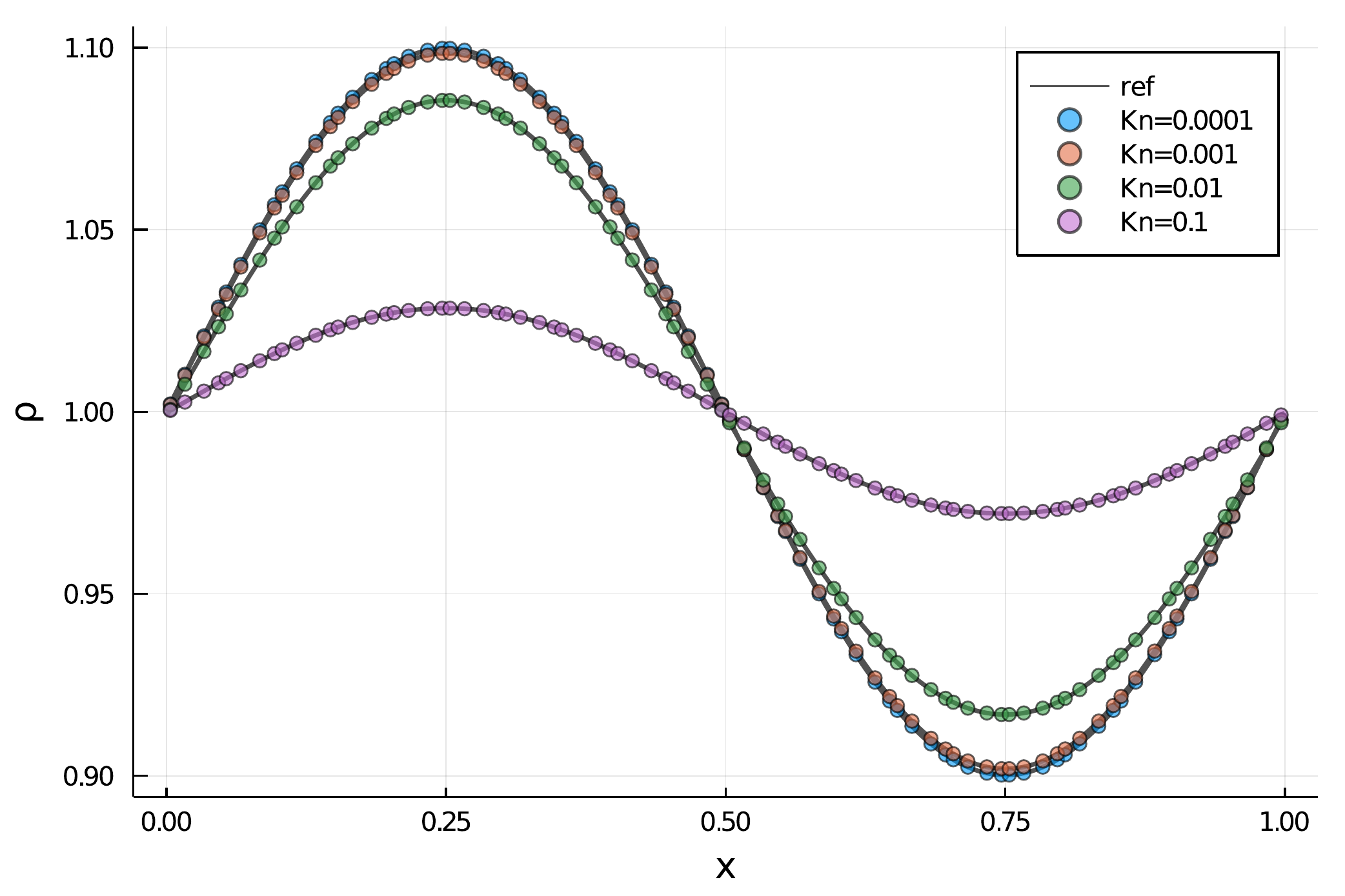}
    \caption{Travelling wave solutions with $N_x=20$ and polynomial degree $3$ at different reference Knudsen numbers.}
    \label{fig:wave}
\end{figure}

As listed, the Lagrange polynomials of degree $2$ and $3$ are used in the computation, resulting in third and fourth order of accuracy by design.
Fig. \ref{fig:wave} shows the traveling wave solutions with $N_x=20$ and polynomial degree $3$ at different reference Knudsen numbers.
The reference solutions are produced by the fast spectral method \cite{wu2013deterministic} with 1000 cells.
With the increasing molecular mean free path, the enhanced particle transports result in stronger viscous dissipation and the smeared wave structure.
Table \ref{tab:frks3_4} to \ref{tab:frks4_1} provide the convergence orders of the flux reconstruction kinetic schemes in design of the third (FRKS3) and fourth order of accuracy (FRKS4).
It is clear that the current method preserves the desired accuracy in all Knudsen regimes.


\begin{table}[htbp]
	\caption{Errors and convergences of FRKS3 in the wave propagation problem at $\mathrm{Kn}=0.0001$.} 
	\centering
	\begin{tabular}{lllllll} 
		\hline
		$\Delta x$ & $L^1$ error & Order & $L^2$ error & Order & $L^\infty$ error & Order \\ 
		\hline
		0.2 & 1.688797E-3 &  & 5.120427E-4 & & 2.215626E-4 \\
        0.1 & 2.678810E-4 & 2.66 & 5.438995E-5 & 3.23 & 1.593844E-5 & 3.79 \\
        0.05 & 3.263045E-5 & 3.04 & 4.729391E-6 & 3.52 & 1.014085E-6 & 3.97\\
        0.025 & 4.045788E-6 & 3.01 & 4.135537E-7 & 3.51 & 6.213239E-8 & 4.03\\
		\hline
	\end{tabular} 
	\label{tab:frks3_4}
\end{table}

\begin{table}[htbp]
	\caption{Errors and convergences of FRKS3 in the wave propagation problem at $\mathrm{Kn}=0.001$.} 
	\centering
	\begin{tabular}{lllllll} 
		\hline
		$\Delta x$ & $L^1$ error & Order & $L^2$ error & Order & $L^\infty$ error & Order \\ 
		\hline
		0.2 & 1.657782E-3 &  & 4.984405E-4 & & 2.105969E-4 \\
        0.1 & 2.105969E-4 & 2.98 & 2.105969-5 & 4.56 & 1.499155E-5 & 3.81 \\
        0.05 & 2.621523E-5 & 3.01 & 4.315510E-6 & 2.29 & 9.015856E-7 & 4.06\\
        0.025 & 3.276879-6 & 3.00 & 3.657462E-7 & 3.56 & 5.422131E-8 & 4.06\\
		\hline
	\end{tabular} 
	\label{tab:frks3_3}
\end{table}

\begin{table}[htbp]
	\caption{Errors and convergences of FRKS3 in the wave propagation problem at $\mathrm{Kn}=0.01$.} 
	\centering
	\begin{tabular}{lllllll} 
		\hline
		$\Delta x$ & $L^1$ error & Order & $L^2$ error & Order & $L^\infty$ error & Order \\ 
		\hline
		0.2 & 1.419118E-3 &  & 4.085279E-4 & & 1.671561E-4 \\
        0.1 & 1.785668E-4 & 2.99 & 3.648516E-5 & 3.49 & 1.069315E-5 & 3.97 \\
        0.05 & 2.141419E-5 & 3.06 & 3.119967E-6 & 3.55 & 6.628941E-7 & 4.01\\
        0.025 & 2.579123E-6 & 3.05 & 2.655507E-7 & 3.55 & 4.047194E-8 & 4.03\\
		\hline
	\end{tabular} 
	\label{tab:frks3_2}
\end{table}

\begin{table}[htbp]
	\caption{Errors and convergences of FRKS3 in the wave propagation problem at $\mathrm{Kn}=0.1$.} 
	\centering
	\begin{tabular}{lllllll} 
		\hline
		$\Delta x$ & $L^1$ error & Order & $L^2$ error & Order & $L^\infty$ error & Order \\ 
		\hline
		0.2 & 3.939578E-4 &  & 1.146447E-4 & & 4.812587E-5 \\
        0.1 & 4.804452E-5 & 3.04 & 9.801638E-6 & 3.55 & 2.943710E-6 & 4.03 \\
        0.05 & 5.964627E-6 & 3.01 & 8.636624E-7 & 3.50 & 1.827464E-7 & 4.01\\
        0.025 & 7.427253E-7 & 3.01 & 7.592175E-8 & 3.51 & 1.140249E-8 & 4.00\\
		\hline
	\end{tabular} 
	\label{tab:frks3_1}
\end{table}


\begin{table}[htbp]
	\caption{Errors and convergences of FRKS4 in the wave propagation problem at $\mathrm{Kn}=0.0001$.} 
	\centering
	\begin{tabular}{lllllll} 
		\hline
		$\Delta x$ & $L^1$ error & Order & $L^2$ error & Order & $L^\infty$ error & Order \\ 
		\hline
		0.2 & 1.677265E-4 &  & 4.413744E-5 & & 1.877038E-5 \\
        0.1 & 1.027635E-5 & 4.03 & 1.826210E-6 & 4.60 & 4.743682E-7 & 5.31 \\
        0.05 & 6.395205E-7 & 4.01 & 8.050237E-8 & 4.50 & 1.508162E-8 & 4.98\\
        0.025 & 4.122186E-8 & 3.96 & 3.654658E-9 & 4.46 & 4.724505E-10 & 5.00\\
		\hline
	\end{tabular} 
	\label{tab:frks4_4}
\end{table}

\begin{table}[htbp]
	\caption{Errors and convergences of FRKS4 in the wave propagation problem at $\mathrm{Kn}=0.001$.} 
	\centering
	\begin{tabular}{lllllll} 
		\hline
		$\Delta x$ & $L^1$ error & Order & $L^2$ error & Order & $L^\infty$ error & Order \\ 
		\hline
		0.2 & 1.647013E-4 &  & 4.236274E-5 & & 1.668588E-5 \\
        0.1 & 1.013567E-5 & 4.02 & 1.788745-6 & 4.57 & 4.867548E-7 & 5.10 \\
        0.05 & 6.243648E-7 & 4.02 & 7.876072E-8 & 4.51 & 1.501070E-8 & 5.02\\
        0.025 & 3.962609-8 & 3.98 & 3.528171E-9 & 4.48 & 4.704976E-10 & 5.00\\
		\hline
	\end{tabular} 
	\label{tab:frks4_3}
\end{table}

\begin{table}[htbp]
	\caption{Errors and convergences of FRKS4 in the wave propagation problem at $\mathrm{Kn}=0.01$.} 
	\centering
	\begin{tabular}{lllllll} 
		\hline
		$\Delta x$ & $L^1$ error & Order & $L^2$ error & Order & $L^\infty$ error & Order \\ 
		\hline
		0.2 & 1.444726E-4 &  & 3.450912E-5 & & 1.172881E-5 \\
        0.1 & 8.303901E-6 & 4.12 & 1.491799E-6 & 4.53 & 4.560153E-7 & 4.68 \\
        0.05 & 5.107348E-7 & 4.02 & 6.450361E-8 & 4.53 & 1.445434E-8 & 4.98\\
        0.025 & 3.210330E-8 & 3.99 & 2.853924E-9 & 4.50 & 4.355419E-10 & 5.05\\
		\hline
	\end{tabular} 
	\label{tab:frks4_2}
\end{table}

\begin{table}[htbp]
	\caption{Errors and convergences of FRKS4 in the wave propagation problem at $\mathrm{Kn}=0.1$.} 
	\centering
	\begin{tabular}{lllllll} 
		\hline
		$\Delta x$ & $L^1$ error & Order & $L^2$ error & Order & $L^\infty$ error & Order \\ 
		\hline
		0.2 & 3.957461E-5 &  & 9.537192E-6 & & 3.142089E-6 \\
        0.1 & 2.353530E-6 & 4.07 & 4.209624E-7 & 4.50 & 1.209994E-7 & 4.70 \\
        0.05 & 1.470697E-7 & 4.00 & 1.853411E-8 & 4.51 & 3.762843E-9 & 5.01 \\
        0.025 & 9.404308E-9 & 3.97 & 8.354799E-10 & 4.47 & 1.189082E-10 & 4.98 \\
		\hline
	\end{tabular} 
	\label{tab:frks4_1}
\end{table}

\subsection{Normal shock structure}

We continue considering a well-resolved problem, i.e. the normal shock wave structure.
The initial particle distribution function is set as Maxwellian in correspondence with the following macroscopic variables
\begin{equation*}
\left[\begin{array}{c}
\rho \\
U \\
V \\
W \\
T \\
\end{array}\right]_{{t=0,L}}=\left[\begin{array}{c}
\rho_- \\
U_-  \\
0 \\
0 \\
T_- \\
\end{array}\right], \quad
\left[\begin{array}{c}
\rho \\
U \\
V \\
W \\
T \\
\end{array}\right]_{{t=0,R}}=\left[\begin{array}{c}
\rho_+ \\
U_+  \\
0 \\
0 \\
T_+ \\
\end{array}\right].
\end{equation*}
Based on the reference frame of shock wave, the upstream and downstream gases are related with the well-known Rankine-Hugoniot relation,
\begin{equation}
\begin{aligned}
    \frac{\rho_{+}}{\rho_{-}} &=\frac{(\gamma+1) \mathrm{Ma}^{2}}{(\gamma-1) \mathrm{Ma}^{2}+2}, \\
    \frac{U_{+}}{U_{-}} &=\frac{(\gamma-1) \mathrm{Ma}^{2}+2}{(\gamma+1) \mathrm{Ma}^{2}}, \\
    \frac{T_{+}}{T_{-}} &=\frac{\left((\gamma-1) \mathrm{Ma}^{2}+2\right)\left(2 \gamma \mathrm{Ma}^{2}-\gamma+1\right)}{(\gamma+1)^{2} \mathrm{Ma}^{2}},
\end{aligned}
\end{equation}
where $\rm Ma$ is the upstream Mach number, and $\gamma=5/3$ is the specific heat ratio of monatomic molecule.
The reference state is set with the upstream flow conditions.
As the shock profile is resolved in this case, no stiffness will be introduced and thus we employ the Bogacki-Shampine integrator, which is a third-order explicit Runge-Kutta method \cite{bogacki19893}.
The detailed computation setup is presented in Table \ref{tab:shock}.

\begin{figure}[htb!]
	\centering
	\subfigure[$\mathrm{Ma}=2$]{
		\includegraphics[width=0.47\textwidth]{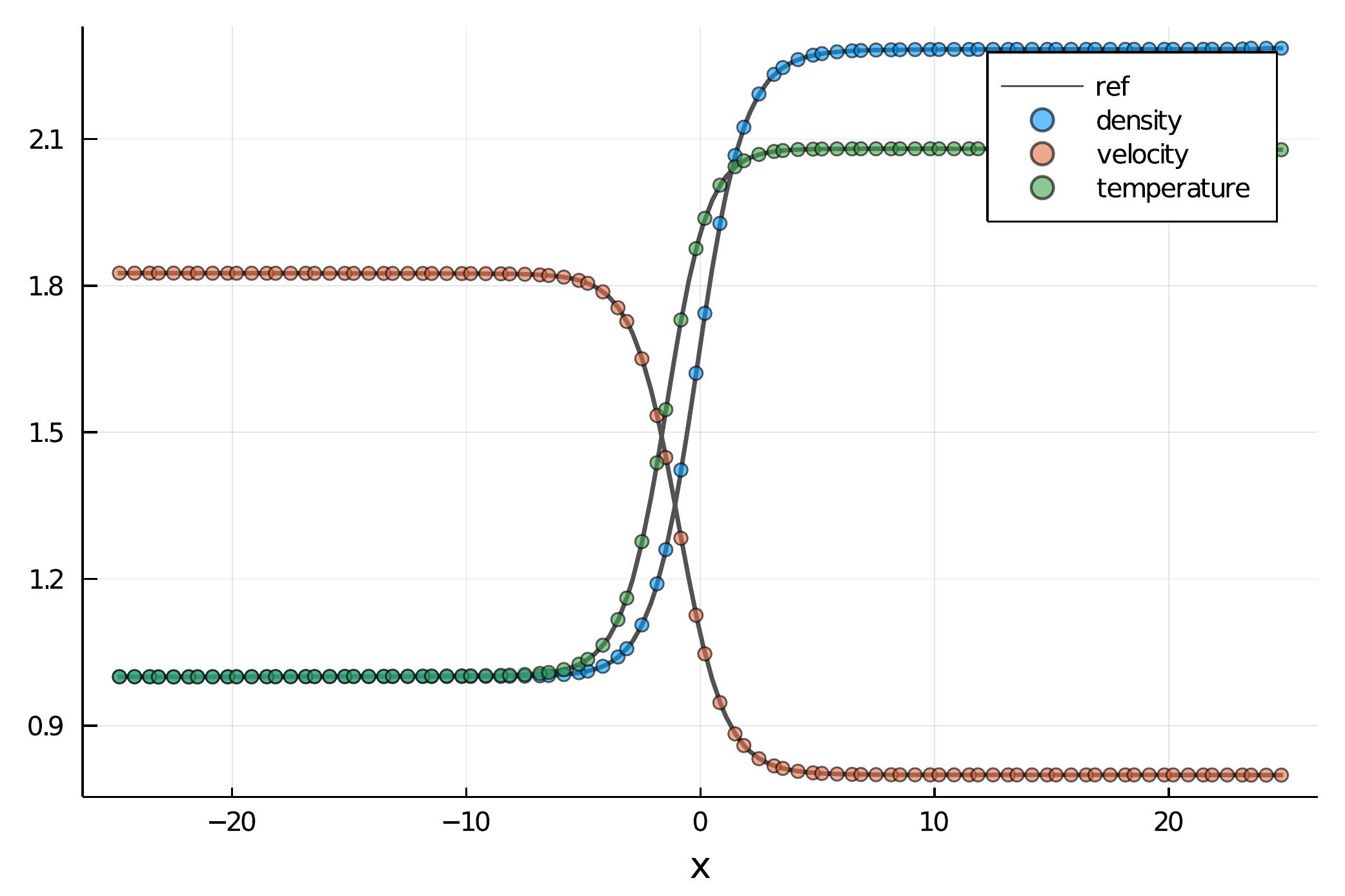}
	}
	\subfigure[$\mathrm{Ma}=3$]{
		\includegraphics[width=0.47\textwidth]{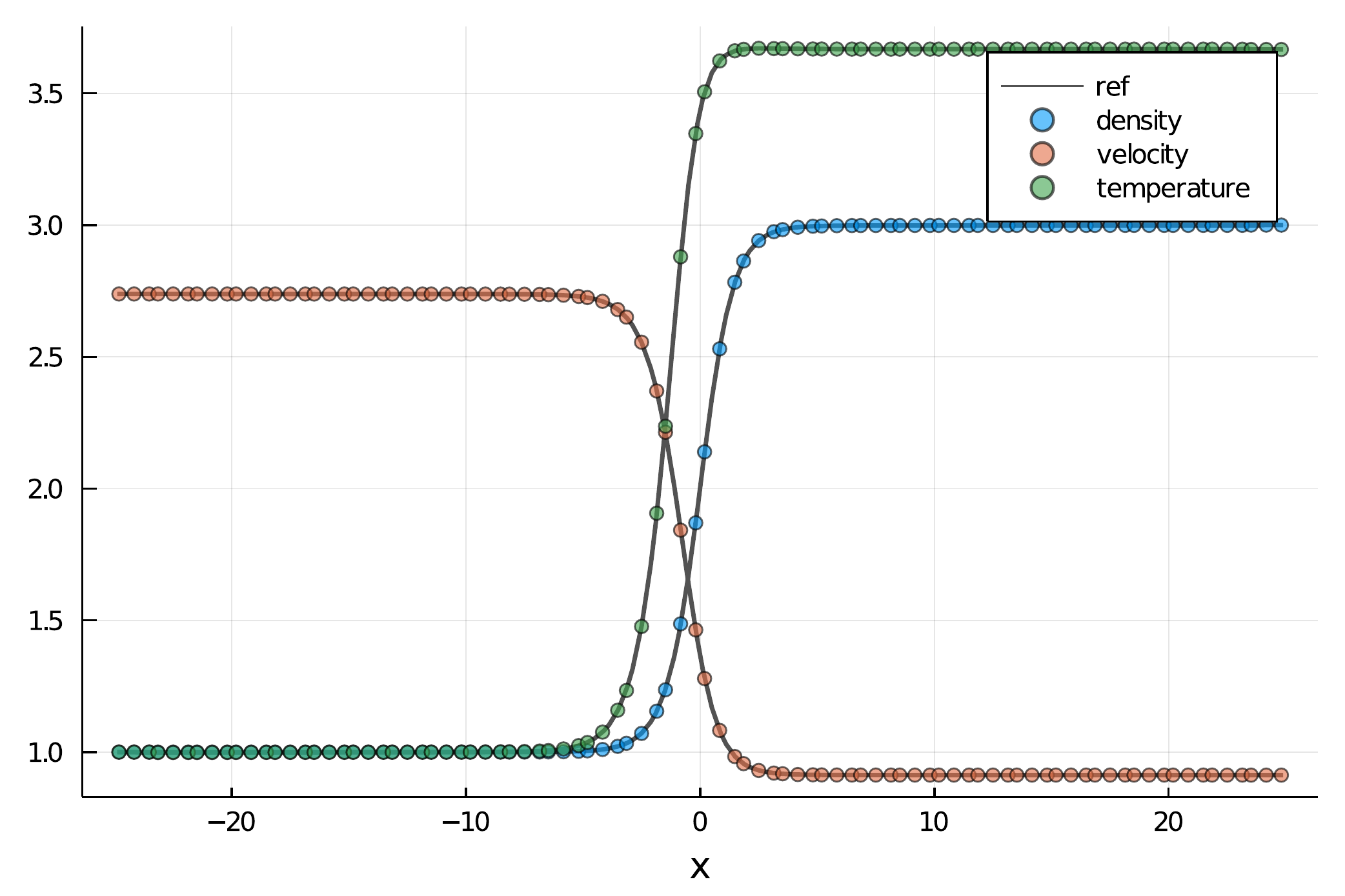}
	}
	\caption{Profiles of density, $U$-velocity and temperature across normal shock wave at different upstream Mach numbers.}
    \label{fig:shock}
\end{figure}

Fig. \ref{fig:shock} provides the profiles of density, $U$-velocity and temperature at different upstream Mach numbers.
The reference solutions are produced by the fast spectral method \cite{wu2013deterministic} with 200 cells.
As is shown, excellent agreement has been achieved between the flux reconstruction solutions under a coarse mesh and the reference results.
It demonstrates the capability of the current scheme to simulate the evolution of non-equilibrium particle distributions.
\begin{table}[htbp]
	\caption{Computational setup of normal shock structure.} 
	\centering
	\begin{tabular}{lllllll} 
		\hline
		$x$ & $N_x$ & Polynomial & Degree & Points & Correction \\ 
		\hline 
		$[-25,25]$ & $50$ & Lagrange & $2$ & Legendre & Radau \\ 
		\hline 
		$\mathbf v$ & $N_u$ & $N_v$ & $N_w$ & Quadrature & Kn  \\ 
		\hline 
		$[-14,14]^3$ & 64 & 32 & 32 & Rectangular & $1.0$  \\ 
		\hline
		Ma & Integrator & Boundary & CFL & \\
		\hline
		$[2,3]$ & Bogacki–Shampine & Dirichlet & 0.2 & \\
		\hline
	\end{tabular} 
	\label{tab:shock}
\end{table}

\subsection{Riemann problem}

Now we shift our gaze to the problem where resolved and unresolved regions coexist in the solution domain.
We employ the Sod shock tube problem, which is a standard one-dimensional Riemann problem.
The particle distribution function is initialized as Maxwellian, which corresponds to the following macroscopic variables
\begin{equation*}
\left[\begin{array}{c}
\rho \\
U \\
V \\
W \\
p \\
\end{array}\right]_{{t=0,L}}=\left[\begin{array}{c}
1 \\
0  \\
0 \\
0 \\
0.5 \\
\end{array}\right], \quad
\left[\begin{array}{c}
\rho \\
U \\
V \\
W \\
p \\
\end{array}\right]_{{t=0,R}}=\left[\begin{array}{c}
0.125 \\
0  \\
0 \\
0 \\
0.1 \\
\end{array}\right].
\end{equation*}
To test the capability of the current scheme to solve resolved/unresolved wave structures and the corresponding multi-scale performance, simulations are performed with different reference Knudsen numbers $\mathrm{Kn}=[0.0001,1]$, with respect to typical continuum, transition, and free molecular flow regimes.
The detailed computation setup is listed in Table \ref{tab:sod}.
\begin{table}[htbp]
	\caption{Computational setup of Sod shock tube.} 
	\centering
	\begin{tabular}{lllllll} 
		\hline
		$t$ & $x$ & $N_x$ & Polynomial & Degree & Points & Correction \\ 
		\hline 
		$[0,0.15]$ & $[0,1]$ & $50$ & Lagrange & $2$ & Legendre & Radau \\ 
		\hline 
		$\mathbf v$ & $N_u$ & $N_v$ & $N_w$ & Quadrature & Kn & CFL \\ 
		\hline 
		$[-8,8]^3$ & 64 & 32 & 32 & Rectangular & $[0.0001,1]$ & $0.15$ \\ 
		\hline
		Integrator & Boundary  \\
		\hline
		ESDIRK-3 & Dirichlet  \\
		\hline
	\end{tabular} 
	\label{tab:sod}
\end{table}

\begin{figure}[htb!]
	\centering
	\subfigure[$\mathrm{Kn}=0.0001$]{
		\includegraphics[width=0.47\textwidth]{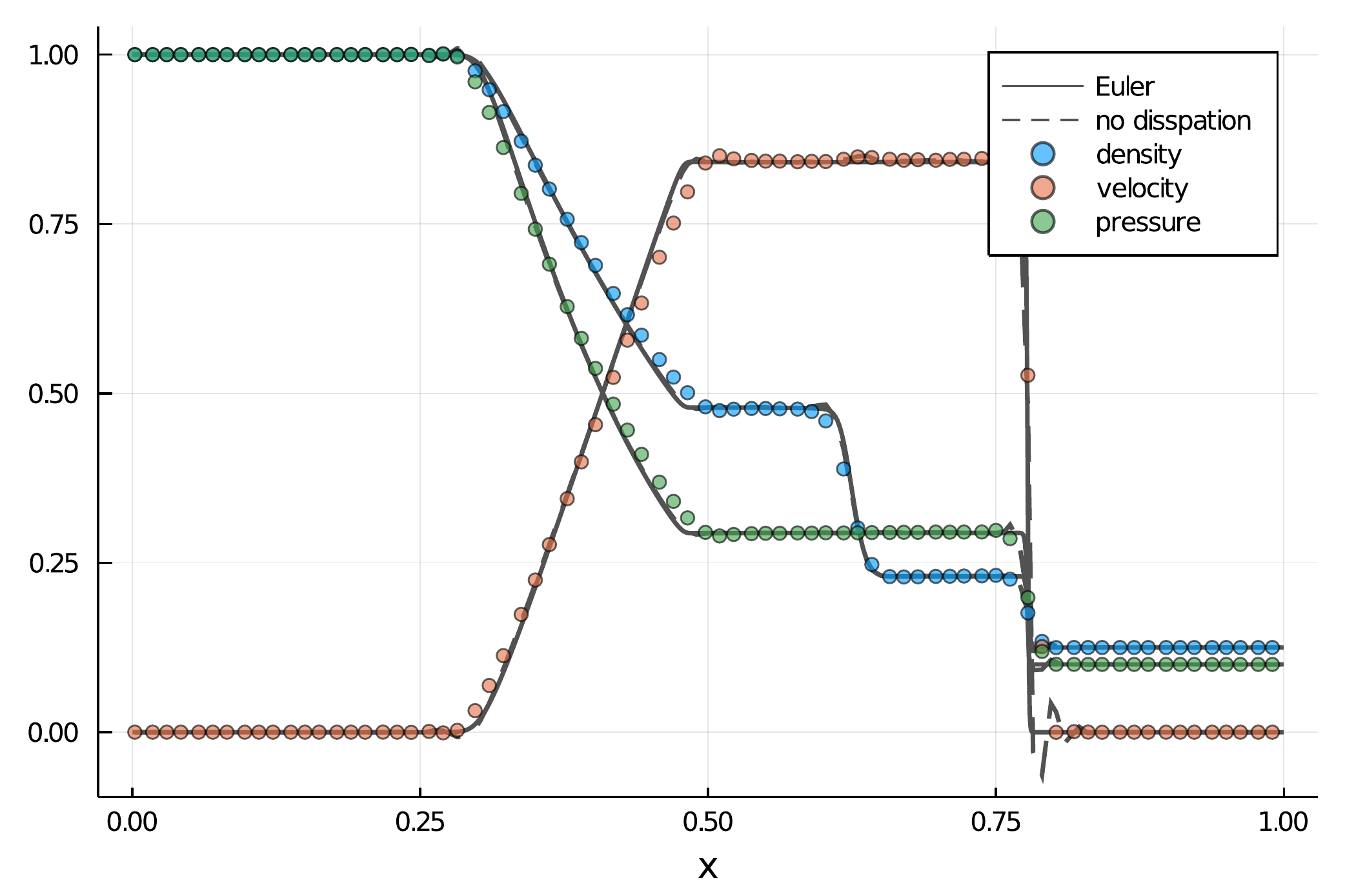}
	}
	\subfigure[$\mathrm{Kn}=0.01$]{
		\includegraphics[width=0.47\textwidth]{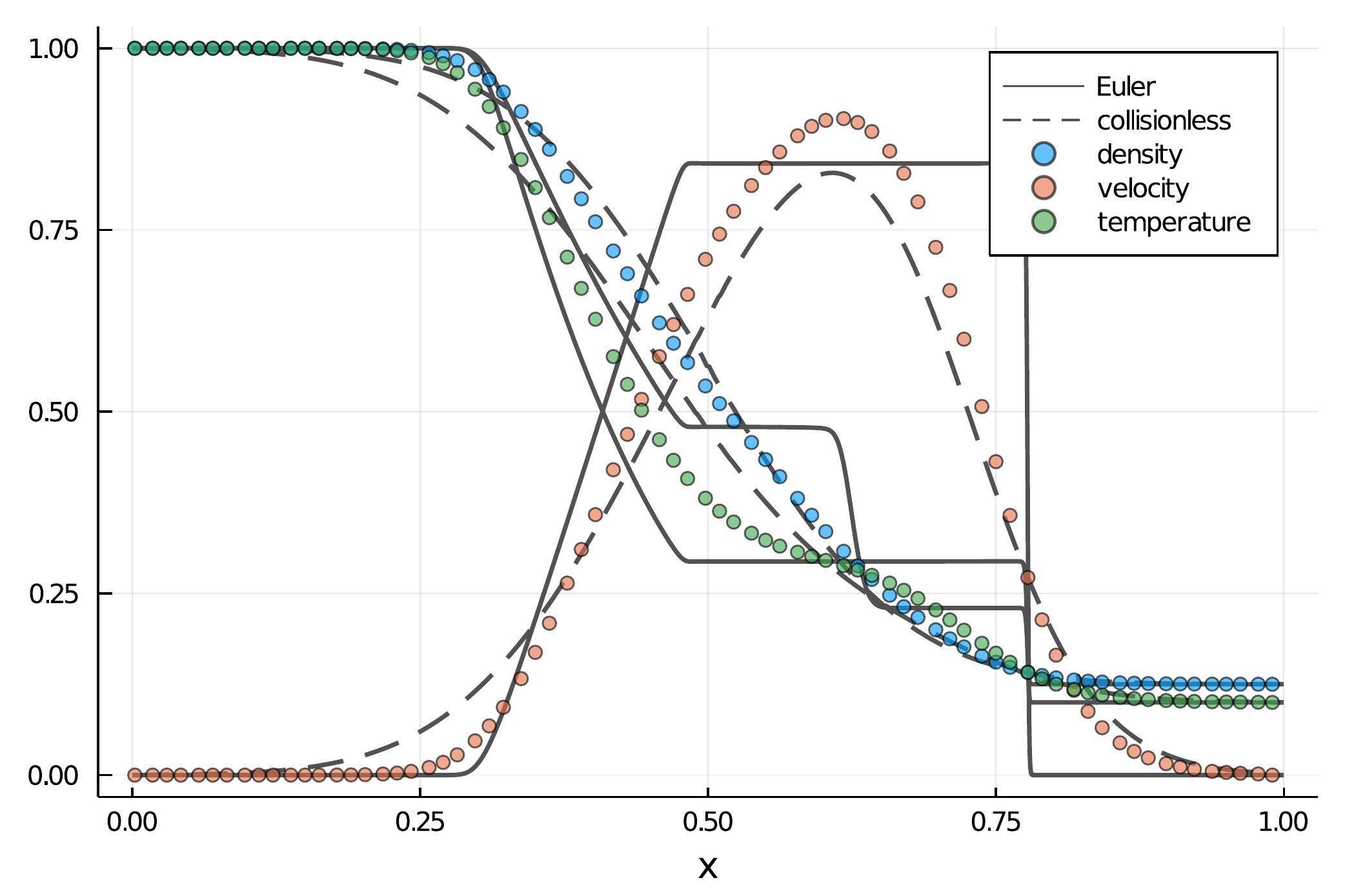}
	}
	\subfigure[$\mathrm{Kn}=1$]{
		\includegraphics[width=0.47\textwidth]{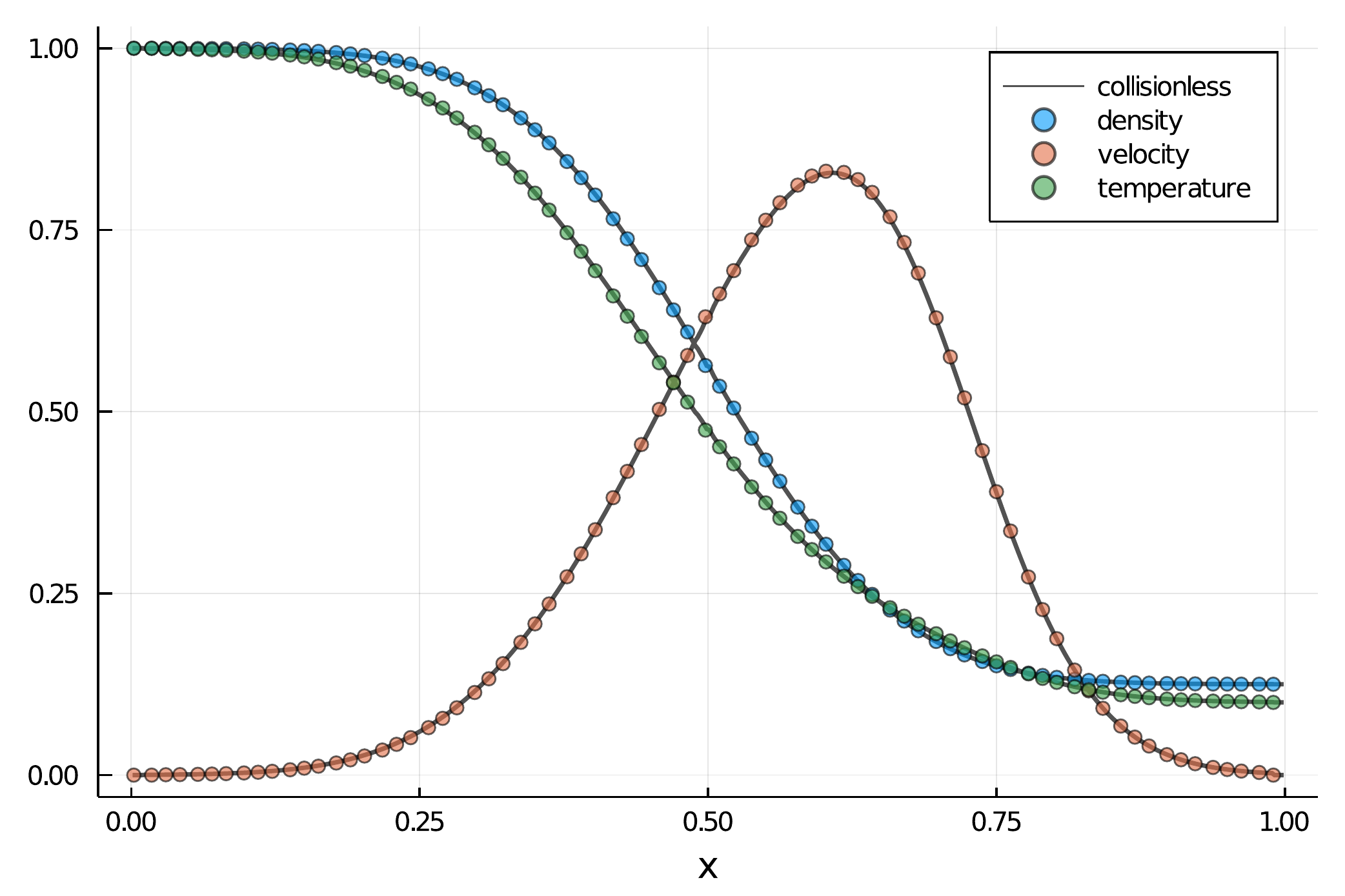}
	}
	\caption{Profiles of density, $U$-velocity and temperature at $t=0.15$ in the Sod shock tube at different reference Knudsen numbers.}
    \label{fig:sod}
\end{figure}

Fig. \ref{fig:sod} presents the profiles of density, $U$-velocity and temperature inside the shock tube at the output instant $t=0.2$ under different Knudsen numbers.
The reference solutions are derived from the Euler and collisionless Boltzmann equations.
In the continuum regime with $\mathrm{Kn}=0.0001$, the molecular mean free path is much less than the grid size, and thus the current method becomes a shock capturing method under limited resolution in space and time.
As shown in Fig. \ref{fig:sod}a, oscillatory solutions from the original flux reconstruction method emerge around the shock wave front due to the Gibbs phenomenon.
Conversely, the adaptive artificial dissipation introduced in the current scheme eliminates the oscillations effectively while preserving the sharp wave structures.
With increasing Knudsen number and molecular mean fee path, the enhanced transport phenomena widen the waves and reduce the gradients of characteristic variables.
Therefore, the solution profiles becomes resolvable under the current resolution.
From $\mathrm{Kn} = 0.0001$ to $\mathrm{Kn} = 1$, the artificial dissipation doesn't destroy the solutions from the current method, and a smooth transition is recovered from the Euler solutions of Riemann problem to the collisionless Boltzmann solutions.

\subsection{Couette flow}

The former cases consider only periodic or Dirichlet boundary conditions.
In this case, we employ the Couette flow as an example to test the gas-surface interactions in the flux reconstruction kinetic scheme.
The initial particle distribution is set as Maxwellian based on the homogeneous fluids,
\begin{equation*}
\left[\begin{array}{c}
\rho \\
U \\
V \\
W \\
T \\
\end{array}\right]_{{t=0}}=\left[\begin{array}{c}
1 \\
0  \\
0 \\
0 \\
1 \\
\end{array}\right].
\end{equation*}
The boundary temperature at both ends of the domain are set as $T_w=1$, and the velocities differ as $\mathbf V_{wL}=[0, -1, 0]^T$, $\mathbf V_{wR}=[0, 1, 0]^T$.
Maxwell's diffusive boundary is adopted to model the gas-surface interaction.
The detailed computational setup is recorded in Table \ref{tab:couette}.
\begin{table}[htbp]
	\caption{Computational setup of Couette flow.} 
	\centering
	\begin{tabular}{lllllll} 
		\hline
		$x$ & $N_x$ & Polynomial & Degree & Points & Correction \\ 
		\hline 
		$[-1,1]$ & $30$ & Lagrange & $2$ & Legendre & Radau \\ 
		\hline 
		$\mathbf v$ & $N_u$ & $N_v$ & $N_w$ & Quadrature & Kn  \\ 
		\hline 
		$[-8,8]^3$ & 72 & 72 & 28 & Rectangular & $[0.2/\sqrt{\pi},20/\sqrt{\pi}]$  \\ 
		\hline
		Integrator & Boundary & CFL & \\
		\hline
		Bogacki–Shampine & Maxwell & 0.15 & \\
		\hline
	\end{tabular} 
	\label{tab:couette}
\end{table}

\begin{figure}
    \centering
    \begin{minipage}{0.49\textwidth}
        \includegraphics[width=\textwidth]{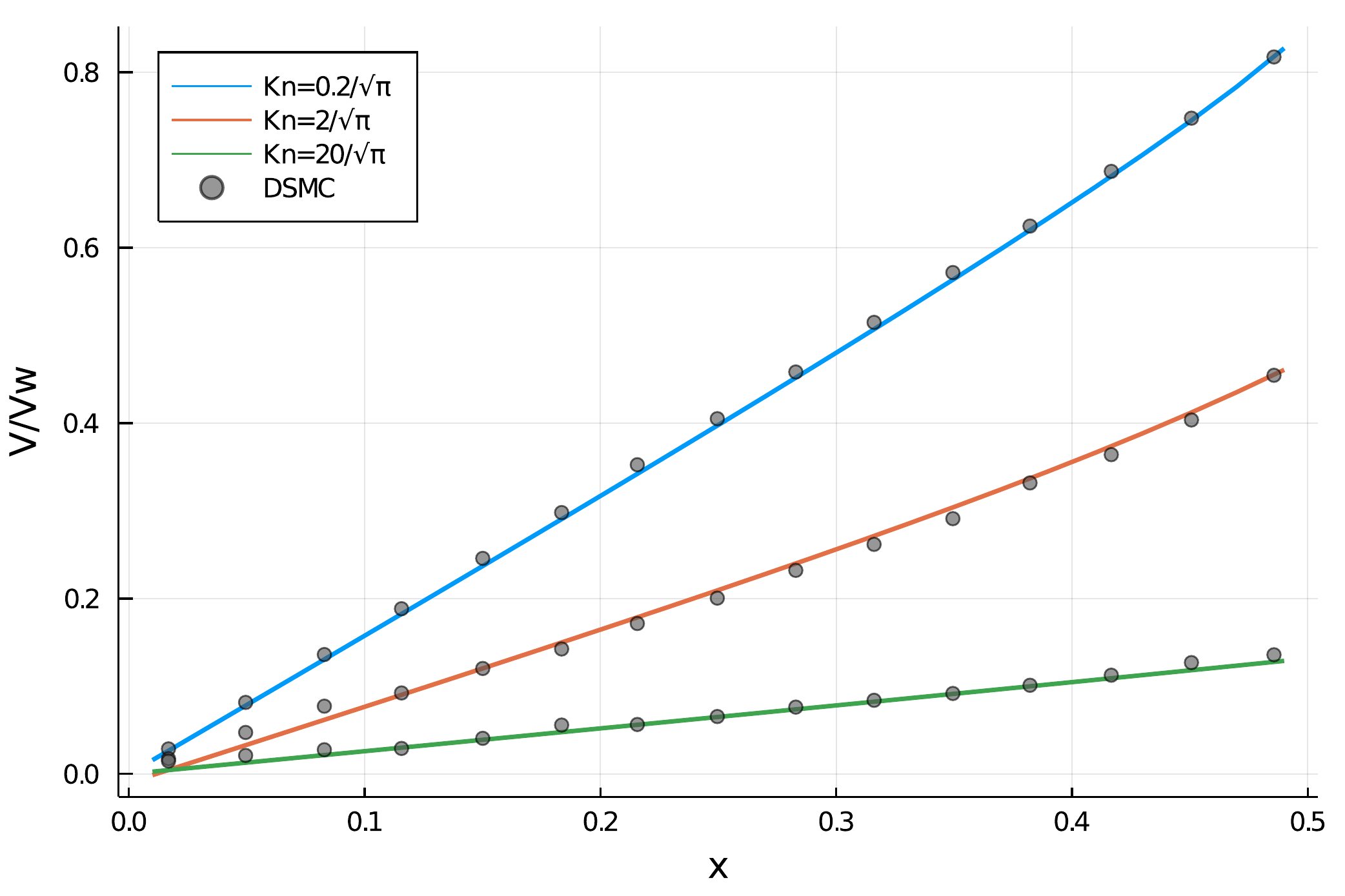}
    \end{minipage}
    \begin{minipage}{0.49\textwidth}
        \includegraphics[width=\textwidth]{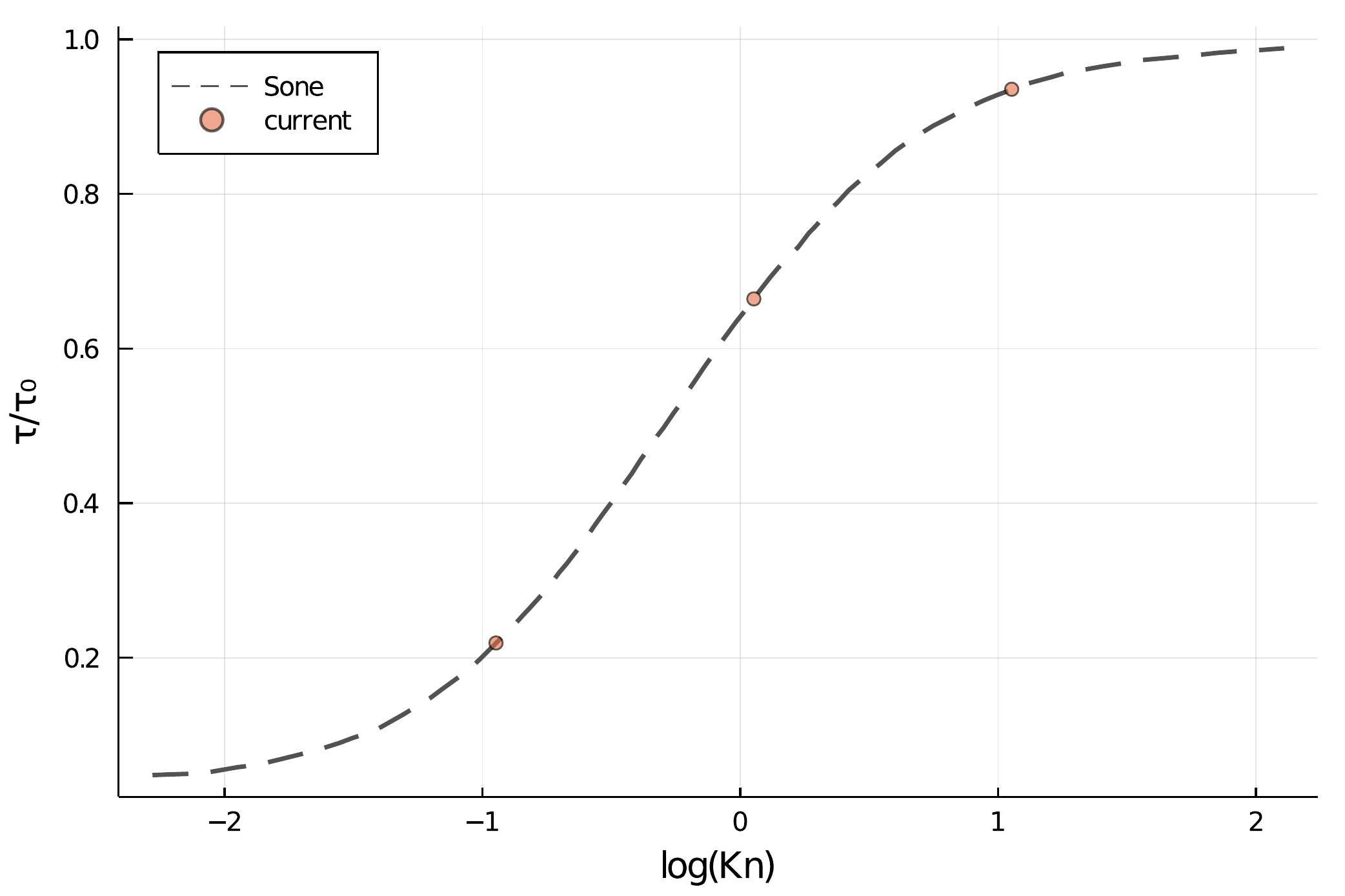}
    \end{minipage}
    \caption{Profiles of $V$-velocity and surface shear stress in the Couette flow at different reference Knudsen numbers.}
    \label{fig:couette}
\end{figure}

Fig. \ref{fig:couette}a shows the macroscopic $V$-velocity profiles in the transition regimes with three Knudsen numbers $\mathrm{Kn}=\{0.2/\sqrt{\pi},2/\sqrt{\pi},20/\sqrt{\pi}\}$.
The current numerical solutions agree perfectly with the reference solutions, which are produced by the information-preserving DSMC method \cite{fan2001statistical}.
Fig. \ref{fig:couette}b draws the relation of surface shear stress versus Knudsen number,
where the collisionless solution is used to determine the normalization factor $\tau_0$.
It is clear that the current solutions fall exactly on the linearized Boltzmann solutions \cite{sone1990numerical} across different Knudsen regimes.

\subsection{Lid-driven cavity}

In the last case, we test the current scheme with multi-dimensional geometry.
The lid-driven cavity is employed as the test problem.
The initial particle distribution is set as Maxwellian with the homogeneous fluids,
\begin{equation*}
\left[\begin{array}{c}
\rho \\
U \\
V \\
W \\
T \\
\end{array}\right]_{{t=0}}=\left[\begin{array}{c}
1 \\
0  \\
0 \\
0 \\
1 \\
\end{array}\right].
\end{equation*}
The solution domain is enclosed by four solid walls with $T_w=1$.
The upper wall moves in the tangent direction with $\mathbf V_w=[0.15, 0, 0]^T$, and the rest three walls are kept still.
Maxwell's diffusive boundary is adopted to all the walls.
The detailed computational setup is provided in Table \ref{tab:cavity}.
\begin{table}[htbp]
	\caption{Computational setup of lid-driven cavity.} 
	\centering
	\begin{tabular}{lllllll} 
		\hline
		$x$ & $y$ & $N_x$ & $N_x$ & Polynomial & Degree \\ 
		\hline 
		$[0,1]$ & $[0,1]$ & $15$ & 15 & Lagrange & $2$ \\ 
		\hline 
		Points & Correction & $\mathbf v$ & $N_u$ & $N_v$ & $N_w$ \\ 
		\hline 
		Legendre & Radau & $[-8,8]^3$ & 32 & 32 & 28 \\ 
		\hline
		Quadrature & Kn & Integrator & Boundary & CFL \\
		\hline
		 Rectangular & $[0.2/\sqrt{\pi},20/\sqrt{\pi}]$ & Bogacki–Shampine & Maxwell & $0.15$ \\
		\hline
	\end{tabular} 
	\label{tab:cavity}
\end{table}

\begin{figure}
    \centering
    \begin{minipage}{0.49\textwidth}
        \includegraphics[width=\textwidth]{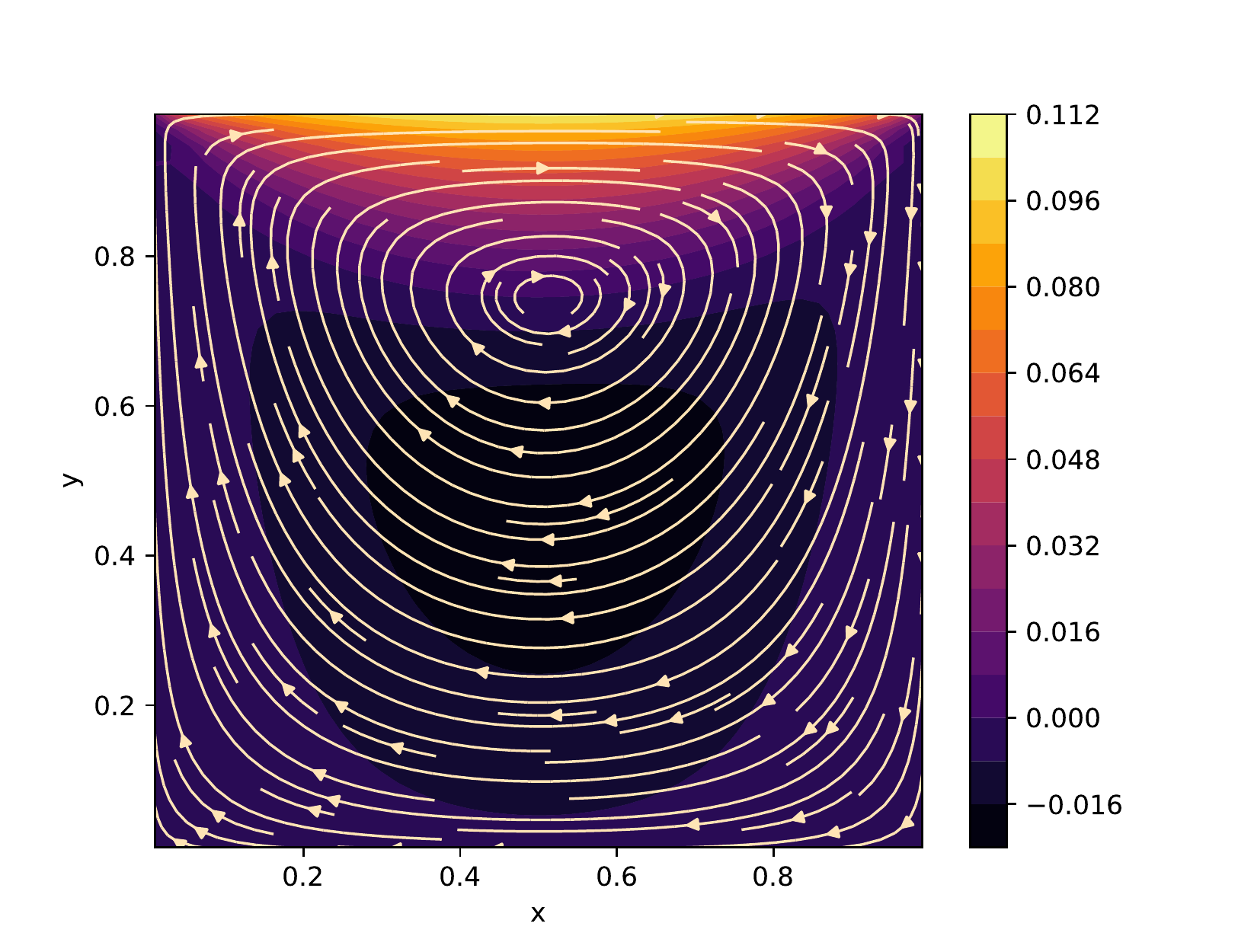}
    \end{minipage}
    \begin{minipage}{0.49\textwidth}
        \includegraphics[width=\textwidth]{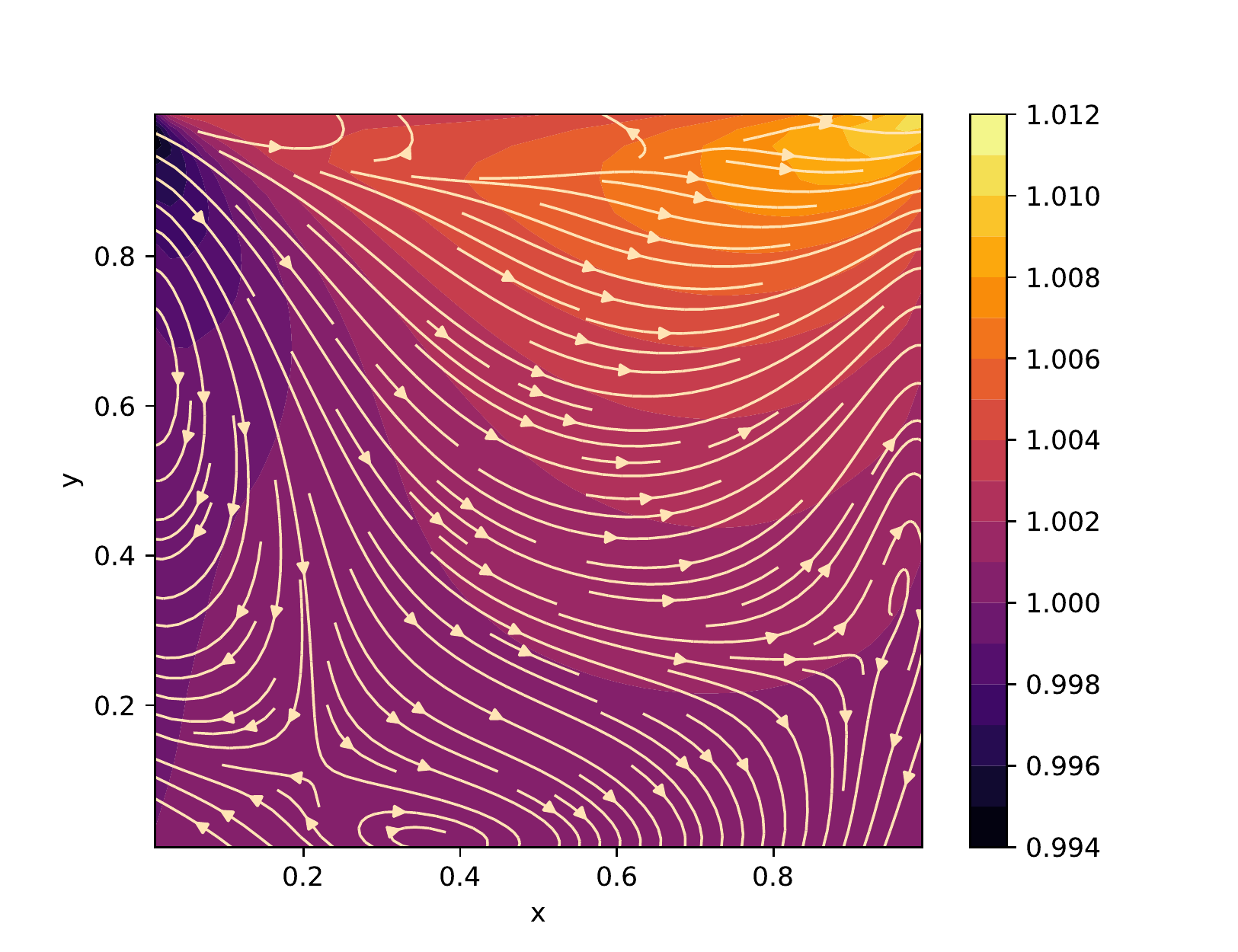}
    \end{minipage}
    \caption{Contours of $U$-velocity with streamlines and temperature with heat flux vectors inside the cavity.}
    \label{fig:cavity contour}
\end{figure}

\begin{figure}
    \centering
    \begin{minipage}{0.47\textwidth}
        \includegraphics[width=\textwidth]{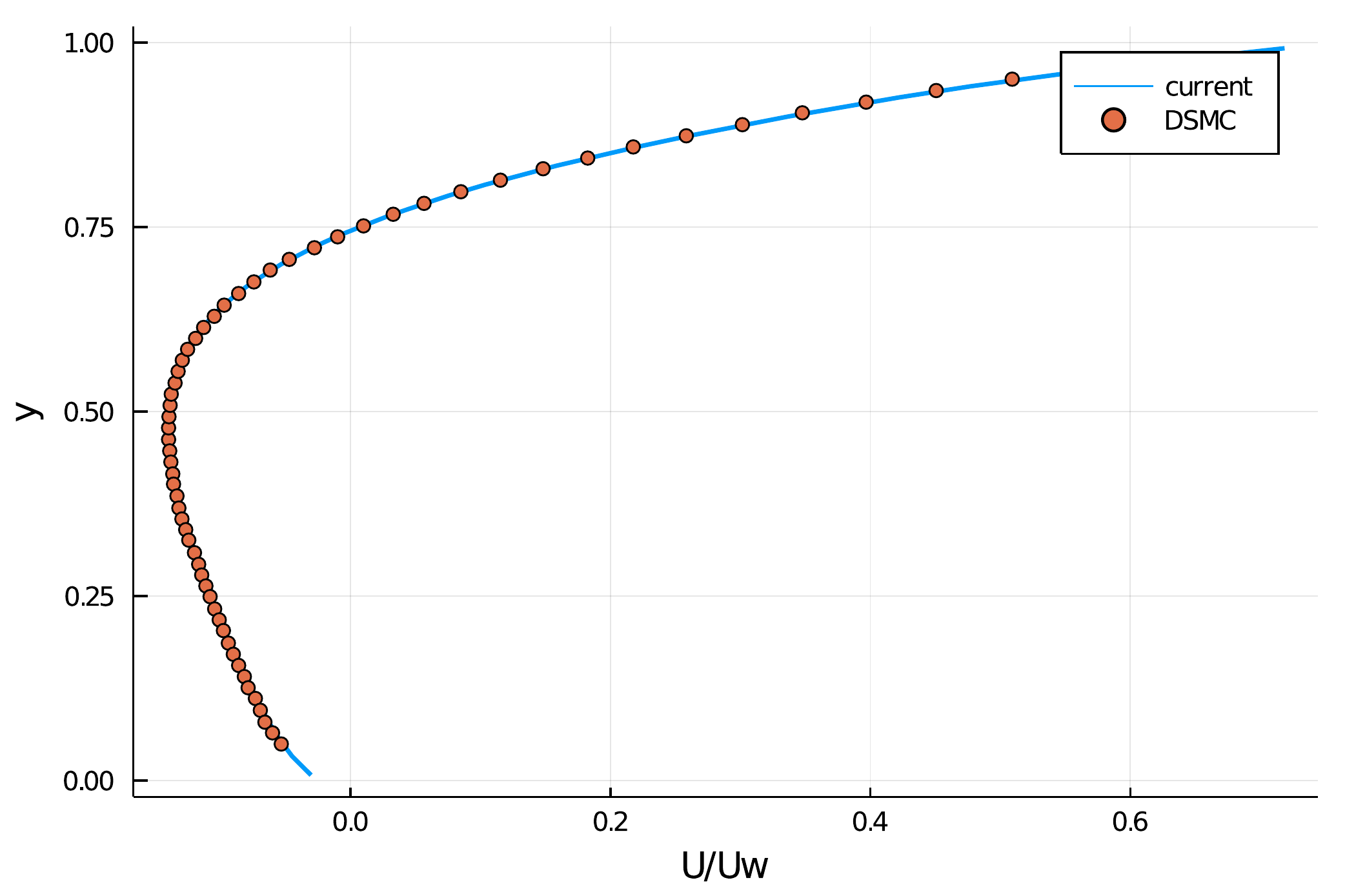}
    \end{minipage}
    \begin{minipage}{0.47\textwidth}
        \includegraphics[width=\textwidth]{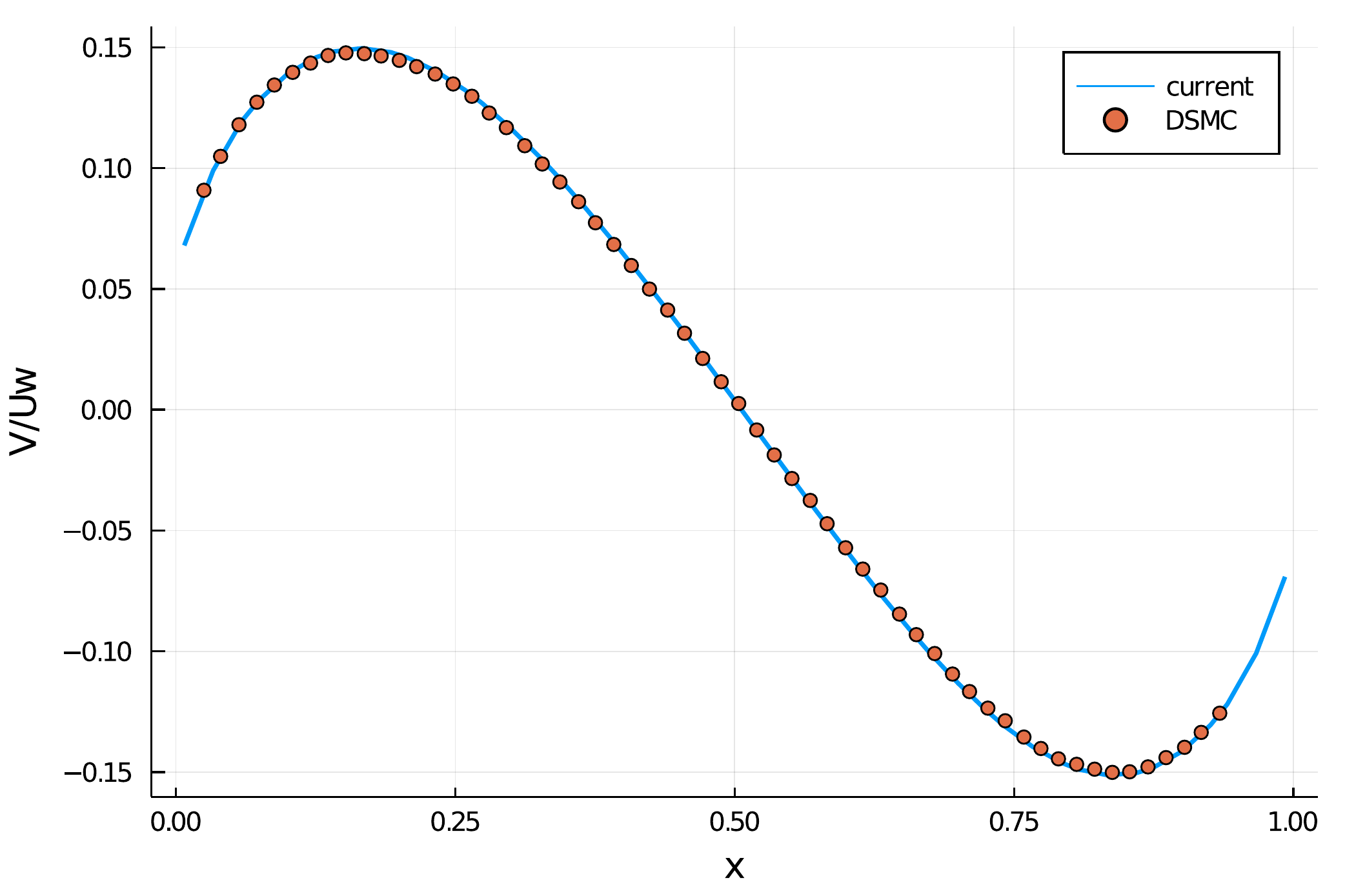}
    \end{minipage}
    \caption{Velocity profiles along vertical and horizontal central lines inside the cavity.}
    \label{fig:cavity line}
\end{figure}

Fig. \ref{fig:cavity contour} shows the contours of $U$-velocity with streamlines and temperature with heat flux vectors inside the cavity.
As explained in \cite{john2011effects}, the anti-Fourier's heat flux driven by stress is clearly identified.
Fig. \ref{fig:cavity line} the velocity profiles along the vertical and horizontal central lines of the cavity.
The DSMC solutions with $60\times 60$ physical mesh are plotted for comparison.
The quantitative comparison demonstrates that the current scheme is able to provide equivalent DSMC solutions in the transition regime with much coarser mesh.

\section{Conclusion}

Non-equilibrium statistical mechanics is profoundly built upon the Boltzmann equation.
For the first time, a high-order kinetic scheme based on flux reconstruction is proposed for solving the Boltzmann equation in this paper.
The upwind flux solver is integrated with flux reconstruction formulation seamlessly throughout the phase space.
The fast spectral method is constructed to solve the exact Boltzmann collision integral with an arbitrary collision kernel.
Besides, the explicit singly diagonally implicit Runge-Kutta method ensures the compatible accuracy in time direction and overcome the stiffness of collision term in the continuum flow regime.
The current method provides an accurate and efficient tool for the study of cross-scale and non-equilibrium flow phenomena.
It shows the potential to be extended to other complex systems, e.g. astrophysics \cite{xiao2018investigation}, plasma physics \cite{xiao2021stochastic}, uncertainty quantification \cite{xiao2020stochastic}, etc.

\clearpage
\newpage

\bibliographystyle{unsrt}
\bibliography{main}
\newpage

\end{document}